\documentclass[11pt]{article}
\usepackage{epsfig}
\usepackage{amsfonts}
\usepackage{amsmath}
\usepackage{bbm}
\usepackage{cite}
\allowdisplaybreaks[1]

\input pix.sty

\newcommand{\rO}{r^{ }_0}
\newcommand{\VT}{V^{ }_\rmii{$T$}}

\newcommand{\gammaE}{\gamma} 
\renewcommand{\eq}{eq.~}
\renewcommand{\eqs}{eqs.~}
\renewcommand{\se}{sec.~}
\renewcommand{\ses}{secs.~}
\renewcommand{\fig}{fig.~}
\renewcommand{\figs}{figs.~}
\newcommand{\dd}{\mathrm{d}}
\newcommand{\tinymsbar}{{\overline{\mbox{\tiny\rm{MS}}}}}
\newcommand{\Lambdamsbar}{{\Lambda_\tinymsbar}}
\newcommand{\mD}{m_\rmii{D}}

\newcommand{\Nf}{N_{\rm f}}
\newcommand{\Nc}{N_{\rm c}}
\newcommand{\Ns}{N_{\rm s}}
\newcommand{\Nt}{N^{ }_\tau}
\newcommand{\Tc}{T_{\rm c}}

\newcommand{\rmO}{{\mathcal{O}}}
\newcommand{\bmu}{\bar\mu}
\newcommand{\CA}{\Nc}
\newcommand{\CF}{C_\rmii{F}}
\newcommand{\TF}{T_\rmii{F}}

\def\lsi{\raise0.3ex\hbox{$<$\kern-0.75em\raise-1.1ex\hbox{$\sim$}}}
\def\gsi{\raise0.3ex\hbox{$>$\kern-0.75em\raise-1.1ex\hbox{$\sim$}}}
\newcommand{\lsim}{\mathop{\lsi}}
\newcommand{\gsim}{\mathop{\gsi}}

\newcommand{\nF}{n_\rmii{F}}
\newcommand{\nB}{n_\rmii{B}}
 \renewcommand{\nF}[1]{n_\rmii{F{#1}}}
 \renewcommand{\nB}[1]{n_\rmii{B{#1}}}
\newcommand{\rmii}[1]{{\mbox{\tiny\rm{#1}}}}
\newcommand{\re}{\mathop{\mbox{Re}}}
\newcommand{\im}{\mathop{\mbox{Im}}}

\newcommand{\Tint}[1]{{\hbox{$\sum$}\!\!\!\!\!\!\!\int\,}_{\!\!\!\!\raise-0.9ex\hbox{$\scriptstyle{#1}$}}}
\newcommand{\Tinti}[1]{{{\Sigma}\!\!\!\!\raise0.3ex\hbox{$\int$}_\rmii{${#1}$}}}

\newcommand{\bi}{\begin{itemize}}
\newcommand{\ei}{\end{itemize}}
\newcommand{\hide}[1]{ }

\newcommand{\alphas}{\alpha_\rmi{s}}

\def\TAsc(#1,#2)(#3,#4,#5)%
{\SetWidth{2.0}\CArc(#1,#2)(#3,#4,#5)\SetWidth{1.0}}
\def\Lwidth{3}

\def\TAgl(#1,#2)(#3,#4,#5){\SetWidth{2.0}\PhotonArc(#1,#2)(#3,#4,#5){\Lwidth}%
{6.283 #3 mul 360 div #4 #5 sub #4 #5 sub mul sqrt mul Tdensity mul}%
\SetWidth{1.0}}
\def\TLgl(#1,#2)(#3,#4){\SetWidth{2.0}\Photon(#1,#2)(#3,#4){\Lwidth}
{#1 #3 sub #1 #3 sub mul #2 #4 sub #2 #4 sub mul add sqrt Tdensity mul}%
\SetWidth{1.0}}
\newcommand{\piC}[1]{\;\parbox[c]{40pt}{\begin{picture}(120,60)(0,-20)
\SetWidth{1.0}\SetScale{0.35} #1 \end{picture}}\;}
\def\ConnectedA(#1,#2,#3){\piC{#1(60,-15)(75,34,146) #2(60,75)(75,214,326)%
 #3(60,60)(20,190,350)%
 \GBoxc(0,30)(10,10){1} \GBoxc(120,30)(10,10){1}%
  }}
\def\ConnectedB(#1,#2,#3){\piC{#1(60,-15)(75,34,146) #2(60,75)(75,214,326)%
 #3(60,60)(60,0)%
 \GBoxc(0,30)(10,10){1} \GBoxc(120,30)(10,10){1}%
  }}
\def\ConnectedC(#1,#2){\piC{#1(60,-15)(75,34,146) #2(60,75)(75,214,326)%
 \GBoxc(0,30)(10,10){1} \GBoxc(120,30)(10,10){1}%
  }}
\def\ConnectedD(#1,#2){\piC{#1(60,-15)(75,34,146) #2(60,75)(75,214,326)%
 \GBoxc(0,30)(10,10){1} \GBoxc(120,30)(10,10){1}%
 \SetWidth{2.0} 
 \Line(55,55)(65,65)%
 \Line(55,65)(65,55)
  }}

\makeatletter \@addtoreset{equation}{section} \makeatother
\renewcommand{\theequation}{\arabic{section}.\arabic{equation}}
\makeatletter
\renewcommand\section{\@startsection {section}{1}{\z@}%
                                   {-5.5ex \@plus -1ex \@minus -.2ex}
                                   {2.3ex \@plus.2ex}%
                                   {\normalfont\large\bfseries}}
\renewcommand\subsection{\@startsection{subsection}{2}{\z@}%
                                     {-3.25ex\@plus -1ex \@minus -.2ex}%
                                     {1.5ex \@plus .2ex}%
                                     {\normalfont\normalsize\bfseries}}
\renewcommand\thesection {\@arabic\c@section}
\renewcommand\thesubsection   {\thesection.\@arabic\c@subsection}
\renewcommand{\@seccntformat}[1]{%
\csname the#1\endcsname.\hspace{1.0em}}
\makeatother

\begin{document}

\flushbottom

\begin{titlepage}

\begin{flushright}
BI-TP 2017/12\\ 
November 2017 
\end{flushright}
\begin{centering}

\vspace*{1.0cm}

{\Large{\bf
 Thermal quarkonium physics in the pseudoscalar channel 
}} 

\vspace{0.8cm}

Y.~Burnier$^{\rm a}$,       
H.-T.~Ding$^{\rm b}$,       
O.~Kaczmarek$^{\rm b,c}$,     
A.-L.~Kruse$^{\rm c}$,   \\[1mm]   
M.~Laine$^{\rm d}$,         
H.~Ohno$^{\rm e,f}$,          
H.~Sandmeyer$^{\rm c}$      

\vspace{0.8cm}

$^\rmi{a}$%
{\em    
        Gymnase de Renens, 
        Av.\ du Silo 1, 
        CH-1020 Renens, Switzerland\\} 

\vspace{0.3cm}

$^\rmi{b}$%
{\em
        Key Laboratory of Quark \& Lepton Physics (MOE) 
        and Institute of Particle Physics, \\  
        Central China Normal University, 
        Wuhan 430079, China\\
}

\vspace{0.3cm}

$^\rmi{c}$%
{\em
       Fakult\"at f\"ur Physik, Universit\"at Bielefeld, 
       33615 Bielefeld, Germany\\
}
\vspace{0.3cm}

$^\rmi{d}$%
{\em
AEC, ITP,  University of Bern, 
Sidlerstrasse 5, CH-3012 Bern, Switzerland\\}

\vspace*{0.3cm}

$^\rmi{e}$%
{\em
        Center for Computational Sciences, University of Tsukuba, 
        Ibaraki 305-8577, Japan\\
}

\vspace*{0.3cm}

$^\rmi{f}$%
{\em
        Physics Department, Brookhaven National Laboratory, 
        Upton, NY 11973, USA\\
}

\vspace*{0.8cm}

\mbox{\bf Abstract}
 
\end{centering}

\vspace*{0.3cm}
 
\noindent
The pseudoscalar correlator
is an ideal lattice probe for thermal modifications to quarkonium 
spectra, given that it is not compromised by a contribution from 
a large transport peak. We construct a perturbative spectral function 
incorporating resummed thermal effects around the threshold and vacuum 
asymptotics above the threshold, and compare the corresponding  
imaginary-time correlators with continuum-extrapolated lattice data 
for quenched SU(3) at several temperatures. Modest differences 
are observed, which may originate from non-perturbative 
mass shifts or renormalization factors, however no resonance peaks
are needed for describing the quenched lattice data for charmonium
at and above $T\sim 1.1\Tc \sim$ 350 MeV. For comparison, in the 
bottomonium case a good description of the lattice data is 
obtained with a spectral function containing 
a single thermally broadened resonance peak. 

\vfill


\vfill

\end{titlepage}

%
\section{Introduction}

Even though potential models
and sum rules have suggested that charmonia
resonances ($J/\psi, \eta_c$) have dissolved by the time that 
the temperature of a QCD plasma reaches 300 MeV or so
(cf.\ e.g.\ refs.~\cite{satz,yb1,sum1,acl} for overviews), it has been 
difficult to consolidate this picture with direct lattice measurements of 
thermal 2-point correlators corresponding to vector
and pseudoscalar operators. 
A part of the problem is that 
spectral information is well hidden 
in an imaginary-time measurement. For instance, in 
the vector channel ($J/\psi$), 
the imaginary-time correlator gets a substantial
contribution from a ``transport peak'' located at small frequencies 
$|\omega| \ll T^2/M $~\cite{tp,umeda}, where $T$ is the temperature and 
$M$ is a heavy-quark mass. This reflects interesting physics of 
heavy quark diffusion (cf.\ ref.~\cite{db,kappa} and references therein), 
but has little to do with the fate of 
quarkonium resonances at $\omega \sim 2 M$, even though it severely hampers 
the corresponding imaginary-time measurement~\cite{umeda}.\footnote{%
 A way to avoid the problem is to make use of NRQCD or pNRQCD 
 (cf.\ refs.~\cite{aarts,akr,bkr1,bkr2,kim} and references therein), 
 however their applicability to the charm quark case is questionable 
 and no strict continuum limit can be taken. 
 }  

In contrast to the vector channel, no transport peak is 
expected to be present 
in the pseudoscalar channel ($\eta_c$)~\cite{pert2,amr}. 
The inertness of
the pseudoscalar correlator then led ref.~\cite{bw} to 
conclude that there is no sign of melting of the 
corresponding charmonium resonances
(another recent analysis can be found in ref.~\cite{ikeda}).
However, another lattice study~\cite{ding2}
did find small changes in the pseudoscalar case 
(in contrast to ref.~\cite{bw}, ref.~\cite{ding2} 
made use of the quenched approximation; 
we return to this in \se\ref{se:concl}). Moreover, 
for the vector channel, it was subsequently shown~\cite{GVtau} that within
statistical and systematic uncertainties the results of ref.~\cite{ding2}
are compatible with perturbative predictions involving 
{\em no} charmonium resonances (only 
a modest threshold enhancement). This demonstrates
that $J/\psi$ and $\eta_c$ dissociation is a possibility and that 
a very fine resolution is necessary for constraining this physics. 

The purpose of the current study is to carry out 
considerations such as those in refs.~\cite{ding2,GVtau} 
but for the pseudoscalar correlator and 
with improved resolution.\footnote{%
 The scalar correlator has also been considered~\cite{ding2,GStau}. 
 It couples to $p$-wave states around the threshold 
 and therefore addresses different physics, however on the formal
 side there are many similarities with the pseudoscalar
 correlator, such as an extreme sensitivity to quark mass definitions
 and the threshold location~\cite{proc}.  
 } 
The pseudoscalar correlator
reflects the physics of the $\eta_c$ but couples to the
same non-relativistic operators 
as the vector channel as far as spin-independent terms
are concerned. 

Our presentation is organized as follows. 
After defining the basic setup in \se\ref{se:setup},
the vacuum asymptotics of the pseudoscalar
spectral function is reviewed in \se\ref{se:vacuum}, whereas 
\se\ref{se:thermal} contains a discussion of 
thermal effects around the quark-antiquark threshold. 
The lattice setup, 
quark-mass interpolation, renormalization, and continuum extrapolation 
are summarized in \se\ref{se:lattice}.
The perturbative and lattice results are compared in 
\se\ref{se:model}, and conclusions and an outlook
are offered in \se\ref{se:concl}.
Two appendices detail the thermal correlator at 
strict next-to-leading order (NLO), 
as a function of imaginary time and real frequency, respectively. 

%
\section{Basic setup}
\la{se:setup}

We consider QCD with 
one heavy quark, of bare mass $M^{ }_\rmii{B}$ 
and $\msbar$ mass
$m(\bmu)$, and $\Nf$ essentially 
massless dynamical quarks (later on we set $\Nf = 0$). 
The pseudoscalar imaginary-time correlator is defined as 
\ba
 G^{ }_\rmii{P}(\tau)
 & \equiv & 
 M_\rmii{B}^2 \int_\vec{x} 
 \Bigl\langle 
 (\bar\psi i \gamma^{ }_5 \psi) (\tau,\vec{x}) 
 \;
 (\bar\psi i \gamma^{ }_5  \psi) (0,\vec{0})
 \Bigr\rangle^{ }_{\rmi{c}}
 \;, \quad
 0 < \tau < \frac{1}{T}
 \;, \la{GPS_def}
\ea
where $\psi$ denotes the heavy quark Dirac spinor, 
and $\langle...\rangle^{ }_\rmi{c}$ indicates that only the 
connected (i.e.\ flavour non-singlet)
contraction is included in the thermal average. 
Defined this way, the correlator is believed to be 
finite after mass and gauge coupling renormalization, 
even though care needs to be taken with the regularization of 
$\gamma^{ }_5$ (see below). 

In the presence of a spatial ultraviolet regulator, the correlator can be 
Fourier-transformed. Denoting by 
$\omega_n$ a bosonic Matsubara frequency, the 
Fourier transform is given by  
$
 \tilde G^{ }_\rmii{P}(\omega_n) \equiv 
 \int_0^{1/T} \! {\rm d}\tau \,  e^{i\omega^{ }_n \tau} 
 G^{ }_\rmii{P}(\tau)
$. 
The corresponding spectral function is formally given by the cut 
\be
 \rho^{ }_\rmii{P}(\omega) = \im \tilde G^{ }_\rmii{P}(-i [\omega + i 0^+])
 \;, 
\ee
and the correlator of \eq\nr{GPS_def} is reproduced through
\be
  G^{ }_\rmii{P}(\tau) \; = \; 
 \int_0^\infty
 \frac{{\rm d}\omega}{\pi} \rho^{ }_\rmii{P} (\omega)
 \frac{\cosh \left( \left(\frac{1}{2T} - \tau\right)\omega \right) }
 {\sinh\left( \frac{\omega}{2 T} \right) } 
 \;. \la{relation}
\ee

If present,
a narrow ``transport peak'', 
 $\rho^{ }_\rmii{P}/\omega \propto \delta(\omega)$, 
would yield a constant ($\tau$-independent) contribution
to $G^{ }_\rmii{P}(\tau)$. 
However, we have checked by an 
explicit computation (cf.\ appendix~A)
that 
$G^{ }_\rmii{P}$ contains no $\tau$-independent 
contribution at NLO. 
This can be partly understood from 
the partially conserved axial current (PCAC) relation 
\be
 \partial_\mu
 \bigl[
   \bar{\psi} \gamma_\mu \gamma^{ }_5 T^a \psi 
 \bigr]
 = 2 M^{ }_\rmii{B}
   \bar{\psi}\,  \gamma^{ }_5 T^a \psi 
 \;, 
 \la{pcac}
\ee
where $T^a$ is a traceless matrix in flavour space. 
According to \eq\nr{pcac} the pseudoscalar correlator 
can be obtained from a corresponding axial charge
correlator through derivatives, 
$
 \partial_\tau^2 G^{ }_{\rmii{A}^{ }_{0}} = 4 G^{ }_\rmii{P}
$,
so that constant parts appearing 
in $ G^{ }_{\rmii{A}^{ }_{0}} $ get deleted. 
However, at finite temperature
the argument is not rigorous, since in principle
$G^{ }_{\rmii{A}^{ }_{0}}$ could contain $\sim \tau(\frac{1}{T}-\tau)$
which would yield a constant $G^{ }_\rmii{P}$ after 
the second derivative. As mentioned, up to NLO no such term is found.  
A physical reason for the absence of a constant part is that 
they only appear in cases in which the operators 
are related to a conserved current in some limit 
(for example, the scalar density equals the fermionic part of the 
``trace anomaly'' ${T^\mu_{ }}^{ }_{\mu}$). 

In order to estimate $\rho^{ }_\rmii{P}$, it is 
useful to have a physical picture in mind. As a cut, 
the spectral function describes  
a hadronic decay width of a pseudoscalar meson. The spatial
average in \eq\nr{GPS_def} implies that the meson is at rest with
respect to the heat bath. The energies of the decay 
products are of the order $\omega/2$. 
For $\omega \gg 2 M \gg \pi T$, 
where $M$ denotes a pole mass, all thermal 
effects are small (exponentially suppressed at leading order, 
power-suppressed in general~\cite{sch}), because thermal motion
represents a minor correction to the
kinematics of the decay products. In this regime,
the spectral function can be extracted from vacuum computations.  
Decreasing the 
energy to $\omega \sim 2 M$, we are entering the threshold region. 
In this situation even the vacuum computations become less precise, 
because the decay products move slowly and have much
time for final-state interactions, leading to the 
so-called Sommerfeld effect
(cf.\ ref.~\cite{fadin-top} and references therein). Once the momenta of the 
decay products are of the same order as the Debye mass, 
$M v \sim \mD^{ } \sim gT$ where $g \equiv \sqrt{4\pi\alphas}$, 
thermal corrections are of order unity~\cite{ms}. 
Below the threshold, thermal corrections represent
the dominant physics: 
the real-time static potential develops an imaginary part~\cite{imV,bbr,jacopo}
which leads to the dissociation of quarkonium resonances if
$T \gg \alphas M$~\cite{peskin,soto,wu}. 

This rich physics implies that a number of different 
techniques are needed for a reasonably
precise estimate of the pseudoscalar
spectral function. The vacuum computations are discussed in 
\se\ref{se:vacuum} and the thermal ones in \se\ref{se:thermal}. 

%
\section{Vacuum contribution above the threshold}
\la{se:vacuum}

We start with a brief non-expert
review of vacuum results for the pseudoscalar spectral 
function, relevant for $M v \gg \mD^{ }$. In this regime thermal effects 
are power-suppressed~\cite{sch}. Many computations have been 
carried out in the ``on-shell'' (i.e.\ ``pole-mass'') scheme,
keeping the gauge coupling in the $\msbar$ scheme.  
Like above, we may then define the pseudoscalar operator as 
$P = M^{ }_\rmii{B} \bar\psi i \gamma^{ }_5 \psi$,\footnote{%
 For the 't Hooft-Veltman definition of $\gamma^{ }_5$, 
 an additional finite renormalization factor needs to 
 be inserted in this relation~\cite{larin}, 
 however for the ``flavour non-singlet'' correlators that 
 we are interested in, ``naive dimensional regularization''
 has been used in the higher-order literature referred to below, 
 and then no additional factor is needed at the present order.
 We return to a discussion of this point below
 \eq\nr{tR2l} and around \eq\nr{larin}.}
and re-expand then $M^{ }_\rmii{B}$ in terms of the 
pole mass $M$~\cite{polemass,mr,kit}. 

The vacuum $\rho^{ }_\rmii{P}$ has been estimated
for a general $\omega/M$ 
up to $\rmO(\alphas^2)$~\cite{cks,ht}, 
and for the asymptotics 
at $\omega \gg M$ and for certain moments 
of the spectral function up to $\rmO(\alphas^3)$~\cite{hmm,kmmm,mm_new}.
The relation of the QCD spectral
function to the corresponding NRQCD one, needed for near-threshold
resummation, is known up to 
$
 \rmO(\alphas^2)
$~\cite{kops}.
Unfortunately these results show poor convergence, 
even for the bottom quark case~\cite{piclum}. 

To illustrate the problem, we write the 
spectral function in the form
\be
 \left. \frac{ \rho^{ }_\rmii{P}(\omega) }{\omega^2 M^2} \right|^\rmi{vac} 
 \; \equiv \; 
 \frac{\Nc}{8\pi}\, R_\rmi{c}^{p}(\omega) 
 \;. \la{def_v}
\ee
The subscript $c$ refers to a connected quark contraction. 
The $R$-function is expanded as 
\ba
 R^{p}_\rmi{c} (\omega) & = &  R^{p(0)}(\omega) + 
 \frac{\alphas(\bmu) }{\pi} \, \CF R^{p(1)}(\omega) 
 \nn 
 & + &  
 \biggl( \frac{\alphas(\bmu)}{\pi} \biggr)^2
 \Bigl[ 
   \CF^2\, R^{p(2)}_A(\omega)
  + \CF \CA\, R^{p(2)}_{N\!A}(\omega)
  + \CF \TF \Nf\, R^{p(2)}_{l}(\omega)
 \Bigr]  
 + \rmO(\alphas^3)
 \;, \hspace*{6mm} \la{Rp}
\ea
where $\CF \equiv (\Nc^2-1)/(2\Nc)$ and $\TF \equiv 1/2$.

The asymptotics of the different contributions read
(cf.\ e.g.\ refs.~\cite{hs,mm}; $\zeta^{ }_n \equiv \zeta(n)$)
\ba
 R^{p(0)}_{ } & \stackrel{\omega \gg M}{\approx} & 
 1  
 \;, \la{Rp0as} \\
 R^{p(1)}_{ } & \stackrel{\omega \gg M}{\approx} & 
   - \fr32 \ln \biggl( \frac{\omega^2}{M^2} \biggr) + \fr94
 \;, \la{Rp1as} \\ 
 R^{p(2)}_{A} & \stackrel{\omega \gg M}{\approx} & 
   \fr98 \ln^2 \biggl( \frac{\omega^2}{M^2} \biggr) 
   - \frac{57}{16} \ln \biggl( \frac{\omega^2}{M^2} \biggr)
    +\frac{109}{32} + 6 (\ln2 -1)\zeta^{ }_2 
   - \frac{15\zeta^{ }_3}{4}
 \;, \la{Rp2A} \\
 R^{p(2)}_{N\!A} & \stackrel{\omega \gg M}{\approx} & 
   - \frac{11}{16} \ln^2 \biggl( \frac{\omega^2}{M^2} \biggr) 
   - \frac{185}{48} \ln \biggl( \frac{\omega^2}{M^2} \biggr)
   - \frac{11}{8} \ln \biggl( \frac{\bmu^2}{\omega^2} \biggr)
                  \ln \biggl( \frac{\omega^2}{M^2} \biggr)
   +  \frac{33}{16} \ln \biggl( \frac{\bmu^2}{\omega^2} \biggr)
 \nn & & \; + \, 
    \frac{49}{6} -3 \Bigl( \ln2 + \fr18 \Bigr)\zeta^{ }_2 
   - \frac{25\zeta^{ }_3}{8}
 \;, \\
 R^{p(2)}_{l} & \stackrel{\omega \gg M}{\approx} & 
    \frac{1}{4} \ln^2 \biggl( \frac{\omega^2}{M^2} \biggr) 
   + \frac{13}{12} \ln \biggl( \frac{\omega^2}{M^2} \biggr)
   + \frac{1}{2} \ln \biggl( \frac{\bmu^2}{\omega^2} \biggr)
                  \ln \biggl( \frac{\omega^2}{M^2} \biggr)
   -  \frac{3}{4} \ln \biggl( \frac{\bmu^2}{\omega^2} \biggr)
 \nn & & \; - \, 
   \frac{31}{12} + \frac{ 3 \zeta^{ }_2}{2} 
   + \zeta^{ }_3
 \;. \la{Rp2as}
\ea
Order by order these describe the full 
$\omega$-dependence~\cite{cks,ht} well already at modest
$\omega \gsim 4 M$, however as a whole the expression does not  
converge. Indeed, even if we choose 
\be
 \bmu \equiv \max( \pi T, \omega)
 \la{bmu}
\ee
for evaluating $\alphas(\bmu)$, so that 
it is small at $\omega \gg M$, its decrease
$\sim 1 / \ln (\omega^2 / \Lambda^2_{\tinymsbar})$ is 
not fast enough to kill the growing $\ln(\omega^2/M^2)$
appearing in the coefficients in \eqs\nr{Rp1as}--\nr{Rp2as}. 

However, it is not necessary to stick to the on-shell scheme. 
The reason for using a pole mass is that
it is formally necessary for defining a
perturbative series around the threshold (cf.\ appendix~B). 
However, the pole mass may be 
re-expanded as an $\msbar$ mass~\cite{polemass,mr,kit}:
\ba
 \frac{m(\bmu)}{M} & = &  
 1 + \frac{\alphas(\bmu)\CF}{\pi}
 \biggl[ -1 - \fr34 \ln\biggl( \frac{\bmu^2}{M^2}\biggr) \biggr]
 \nn & 
 + & \biggl(\frac{\alphas(\bmu)}{\pi} \biggr)^2
 \biggl\{
   \CF^2 
 \biggl[ 
   \frac{9}{32} \ln^2 \biggl( \frac{\bmu^2}{M^2} \biggr)
 + \frac{21}{32} \ln \biggl( \frac{\bmu^2}{M^2} \biggr)
 + \frac{7}{128} - \frac{15 \zeta^{ }_2}{8} - \frac{3\zeta^{ }_3}{4} + 
 3 \zeta^{ }_2 \ln(2)
 \biggr]
 \nn & & \; + \, 
   \CF \CA 
 \biggl[
  - \frac{11}{32} \ln^2 \biggl( \frac{\bmu^2}{M^2} \biggr)
 - \frac{185}{96} \ln \biggl( \frac{\bmu^2}{M^2} \biggr)
 - \frac{1111}{384} + \frac{\zeta^{ }_2}{2} + \frac{3\zeta^{ }_3}{8} -
 \frac{3 \zeta^{ }_2}{2} \ln(2)
 \biggr]
 \nn & & \; + \, 
   \CF \TF \Nf  
 \biggl[
  \frac{1}{8} \ln^2 \biggl( \frac{\bmu^2}{M^2} \biggr)
 + \frac{13}{24} \ln \biggl( \frac{\bmu^2}{M^2} \biggr)
 + \frac{71}{96} +\frac{ \zeta^{ }_2 }{2} 
 \biggr]
 \biggr\}
 + \rmO(\alphas^3)
 \;. \la{polemass}
\ea
Changing the normalization from \eq\nr{def_v} to
\be
 \left. \frac{ \rho^{ }_\rmii{P}(\omega) }
 {\omega^2 m^2(\bmu)} \right|^\rmi{vac} 
 \; \equiv \; 
 \frac{\Nc}{8\pi}\, \tilde{R}_\rmi{c}^{p} (\omega,\bmu)
 \la{tilde_def_v}
\ee
and writing $\tilde{R}_\rmi{c}^{p}$ as in 
\eq\nr{Rp}, mass dependence drops out from the asymptotics: 
\ba
 \tilde{R}^{p(0)}_{ } & \stackrel{\omega \gg m^{ }(\bmu) }{\approx} & 
  1 
 \;, \la{tR0} \\
 \tilde{R}^{p(1)}_{ } & \stackrel{\omega \gg m^{ }(\bmu) }{\approx} & 
   \fr32 \ln \biggl( \frac{\bmu^2}{\omega^2} \biggr) + \fr{17}4
 \;, \\ 
 \tilde{R}^{p(2)}_{A} & \stackrel{\omega \gg m^{ }(\bmu) }{\approx} & 
   \fr98 \ln^2 \biggl( \frac{\bmu^2}{\omega^2} \biggr) 
   + \frac{105}{16} \ln \biggl( \frac{\bmu^2}{\omega^2} \biggr)
    +\frac{691}{64} - \frac{ 9 \zeta^{ }_2}{4}  
   - \frac{9\zeta^{ }_3}{4}
 \;, \\
 \tilde{R}^{p(2)}_{N\!A} & \stackrel{\omega \gg m^{ }(\bmu) }{\approx} & 
    \frac{11}{16} \ln^2 \biggl( \frac{\bmu^2}{\omega^2} \biggr) 
   + \frac{71}{12} \ln \biggl( \frac{\bmu^2}{\omega^2} \biggr)
   + \frac{893}{64} - \frac{11 \zeta^{ }_2}{8} 
   - \frac{31\zeta^{ }_3}{8}
 \;, \\
 \tilde{R}^{p(2)}_{l} & \stackrel{\omega \gg m^{ }(\bmu) }{\approx} & 
   - \frac{1}{4} \ln^2 \biggl( \frac{\bmu^2}{\omega^2} \biggr) 
   -  \frac{11}{6} \ln \biggl( \frac{\bmu^2}{\omega^2} \biggr)
   - \frac{65}{16} + \frac{\zeta^{ }_2}{2} 
   + \zeta^{ }_3
 \;. \la{tR2l}
\ea
We now see that the choice in \eq\nr{bmu} removes large logarithms
and leads to increasingly
small corrections as $\omega$ grows. Therefore \eq\nr{tilde_def_v}
yields a reliable prediction for large $\omega$.

%
\begin{table}[t]

\small{
\begin{center}
\begin{tabular}{cccc}
 $\displaystyle
  \frac{m(\bmu^{ }_\rmi{ref} \equiv \mbox{2~GeV})}{\rmi{GeV}}$ &
 $\displaystyle
  \frac{m(\bmu = m)}{\rmi{GeV}}$ &
 $\displaystyle
  \frac{M^{ }_x}{\rmi{GeV}}$ & 
 $\displaystyle
  \alphas(m(\bmu = m))$ \\[3mm]
 \hline 
  1.0 & 1.14 & 1.19(2) & 0.280 \\ 
  2.0 & 2.00 & 2.11(2) & 0.208 \\
  3.0 & 2.83 & 2.99(2) & 0.181 \\ 
  4.0 & 3.64 & 3.84(2) & 0.166 \\ 
  5.0 & 4.43 & 4.67(3) & 0.155 \\  
 \hline 
\end{tabular} 
\end{center}
}

\vspace*{3mm}

\caption[a]{\small
  The quark masses used on the perturbative side of this study 
  ($\Nf = 0$). We have used 5-loop
  running for the $\msbar$ mass $m$~\cite{mass1,mass2} 
  and for $\alphas$~\cite{alph1,alph2}. 
  The pole-like mass $M^{ }_x$ has been defined through 
  \eq\nr{fixmass}, with $x = 4...8$
  yielding the variation shown.
  For scale setting we use 
  $r^{ }_0 \Lambdamsbar = 0.602(48)$~\cite{r0Lambda}
  and $r^{ }_0 = 0.47(1)$fm~\cite{r0fm}, whereby
  $\Lambdamsbar \simeq 253$~MeV
  (a recent analysis yields
  $r^{ }_0 \Lambdamsbar = 0.593(17)$~\cite{r0Lambda_new}). 
  A potential model suggests that $m(\bmu^{ }_\rmi{ref}) = 1$~GeV 
  is close to the value relevant for the charmonium case, 
  $M^{ }_\rmii{1S} \equiv M^{ }_{\eta_c} / 2  \approx 1.49$~GeV,
  whereas $m(\bmu^{ }_\rmi{ref}) = 5$~GeV is close to 
  the value relevant for the bottomonium case,
  $M^{ }_\rmii{1S} \approx 4.66$~GeV, however 
  these expectations are subject to large uncertainties. 
 }
\label{table:masses}
\end{table}
%

We note in passing that 
the asymptotics in \eqs\nr{tR0}--\nr{tR2l} is 
identical to what is obtained for the connected part of the scalar
density correlator~\cite{hs,mm}. 
This demonstrates that the scalar and 
pseudoscalar densities have the same anomalous dimension, 
which in turn is related to the quark mass anomalous dimension, 
so that $m^2(\bmu) \tilde{R}_\rmi{c}^{p} (\omega,\bmu)$
is independent of $\bmu$. 

Given the poor convergence of \eq\nr{polemass}, the question arises
how we should relate $M$ and $m(\bmu)$ to each other. 
Following old tradition, we parametrize
quark masses through their $\msbar$ values at the scale 
$\bmu^{ }_\rmi{ref} \equiv 2$~GeV. The corresponding running mass 
is obtained from 5-loop running~\cite{mass1,mass2}. One way to fix
the value of $M$ is to require
that \eqs\nr{def_v} and \nr{tilde_def_v} agree at a value $\omega = x M$
where we assume both representations 
to be reliable, say $x=4...8$:
\be
 M_x^2 \equiv m^2(\bmu)\,
 \left.  \frac{\tilde{R}^p_\rmi{c}(\omega,\bmu)}{R^p_\rmi{c}(\omega)}
 \right|_{\bmu=\omega,~ \omega = x M^{ }_x}
 \;. \la{fixmass}
\ee
The results obtained at $\rmO(\alphas^2)$
are shown in table~\ref{table:masses}. 
(There are many other pole mass definitions, 
however none of them are quite satisfactory for our purposes.)

%
\section{Thermal contributions around the threshold}
\la{se:thermal}

Just above the threshold
(i.e.\ for $\omega - 2 M \ll M$), 
the frequency $\omega$ can be parametrized through
a relative velocity as $M v^2 \equiv \omega - 2 M $. If $M v \lsim g T$, 
thermal effects are of order unity~\cite{ms}. 
Because the heat bath breaks Lorentz invariance, their 
technical treatment is much harder than that of vacuum 
contributions: within strict perturbation theory, 
only the NLO level has been reached (cf.\ appendix~B). 
Unfortunately the NLO expression turns out to be rather useless
in practice; it indicates that thermal corrections are 
power-suppressed as $\sim \alphas T^2/M^2$, whereas in reality
there are also thermal corrections only suppressed by (non-integer) 
powers of $\alphas$. These originate from soft-gluon mediated effects
(thermal heavy quark mass correction, Debye screening, 
real $2\to2$ scatterings of heavy quarks off plasma particles), 
and require a resummed 
treatment. We follow here the implementation of ref.~\cite{peskin}. 

Before proceeding 
let us stress that, as is characteristic of resummed treatments, 
the approach can be justified theoretically only in a certain
parametric range (see below). In practice it is also used 
somewhat outside of this range, and for this purpose it is 
``enhanced'' by a number of phenomenological ingredients (see below), 
in order to avoid a drastic breakdown in the latter situations. 

Within the non-relativistic regime $v \ll 1$, the pseudoscalar 
spectral function is related to the vector channel one: 
\be
 \rho^\rmii{NRQCD}_\rmii{P} = 
 \frac{M^2}{3}
 \rho^\rmii{NRQCD}_\rmii{V} 
 \;, \quad
 \omega \approx 2 M
 \;.  \la{factor2}
\ee
For the vector channel, the spectral function is obtained from
a Wightman function $C^{ }_{>}$ as 
\be
 \rho^\rmii{NRQCD}_\rmii{V}(\omega) = 
 \fr12 \Bigl( 1 - e^{-\frac{\omega}{T }}\Bigr)
 \int_{-\infty}^{\infty} \! {\rm d} t \, e^{i \omega t}
 \; C_{>}(t;\vec{0,0})
 \;,  \la{nrqcd_T}
\ee
where $C^{ }_{>}$ is solved from 
\ba
 \biggl\{ i \partial_t - \biggl[ 2 M 
 + \VT(r)
 - \frac{\nabla_\vec{r}^2}{M}
 \biggr] \biggr\} \, C_{>}^V(t;\vec{r,r'})  & =  & 0 
 \;,  \quad t\neq 0 \;, \la{Seq} \\ 
 C_{>}^V(0;\vec{r,r'}) & = & 6 \Nc\, \delta^{(3)}(\vec{r-r'})
 \;. \la{In0}
\ea
For $t > 0$ the potential is of the form~\cite{imV,bbr,jacopo} 
\be
  \VT(r) = 
  -\alphas \CF \biggl[ 
 \mD^{ } + \frac{\exp(-\mD^{ } r)}{r}
 \biggr] - {i \alphas \CF T } \, \phi(\mD^{ } r)  + \rmO(\alphas^2)    
 \;, \la{expl}
\ee
where the function 
\be
 \phi(x) \equiv 
 2 \int_0^\infty \! \frac{{\rm d} z \, z}{(z^2 +1)^2}
 \biggl[
   1 - \frac{\sin(z x)}{zx} 
 \biggr]
 \la{phi}
\ee
represents the effects of real $2\to2$ scatterings 
of the quark and antiquark off medium particles. For $t < 0$
the sign of $\im \VT$ is reversed. 
For numerical estimates we insert 
2-loop values of $\mD^{ }$ and 
a thermal $\alphas$ from ref.~\cite{gE2} 
(for $\mD^{ }$ the 3-loop level has also been 
reached~\cite{mE2}). 

At short separations, $r \ll 1/\mD^{ }$, we replace the thermal potential 
by a vacuum one. At 2-loop level, 
\ba
 V^{ }_0 (r) & = & 
 - \int\! \frac{{\rm d}^3\vec{k}}{(2\pi)^3}
 \biggl( \frac{4\pi \CF\alphas e^{i\vec{k}\cdot\vec{r}}}{k^2} \biggr)
 \biggl\{
   1 + \frac{\alphas}{4\pi}
   \biggl[ a^{ }_1 + \beta^{ }_0 \ln\biggl(\frac{\bmu^2}{k^2}\biggr) \biggr] 
 \nn 
 &  & \hspace*{6mm} + \, \biggl( \frac{\alphas}{4\pi} \biggr)^2
   \biggl[
           a^{ }_2 + \bigl(\beta^{ }_1 + 2 a^{ }_1\beta^{ }_0 \bigr)
           \ln\biggl(\frac{\bmu^2}{k^2}\biggr) 
           + \beta^2_0 \ln^2\biggl( \frac{\bmu^2}{k^2} \biggr)
   \biggr]
 \biggr\}
 + \rmO(\alphas^4)
 \\
 & = & 
 - \frac{\CF\alphas(\frac{e^{-\gammaE}}{r})}{r}
 \biggl\{
    1 +  \frac{\alphas(\frac{e^{-\gammaE}}{r})}{4\pi} \, a^{ }_1
      +  \frac{\alphas^2(\frac{e^{-\gammaE}}{r})}{(4\pi)^2} \,
     \biggl[  a^{ }_2 + \frac{\pi^2 \beta_0^2}{3} \biggr]
 \biggr\}
 + \rmO(\alphas^4)
 \;, \la{V0}
\ea
where $\gammaE$ is the Euler constant and, for $\Nf = 0$~\cite{pot1}, 
\be
 a^{ }_1 = \frac{31\Nc}{9}
 \;, \quad
 a^{ }_2 = \biggl( \frac{4343}{162} + 4\pi^2 - \frac{\pi^4}{4}
 + \frac{22\zeta^{ }_3}{3} \biggr) \Nc^2
 \;, \quad
 \beta^{ }_0 = \frac{11\Nc}{3} 
 \;, \quad
 \beta^{ }_1 = \frac{34\Nc^2}{3} 
 \;.
\ee
The 3-loop potential is also known 
in analytic form by now~\cite{pot4}. 
At $r\mD^{ } \ll 1$ the potential of \eq\nr{V0} is less binding than that
of \eq\nr{expl} because the coupling runs towards zero; at
$r\mD^{ } \gg 1 $ \eq\nr{expl} is less binding because 
of Debye screening. For our numerics, 
we employ $\mbox{max}\{ V^{ }_0,\re[\VT(r) - \VT(\infty) ]\}$
as the $r$-dependent real part of the potential, 
whereby $V^{ }_0$ applies at short separations and 
$\re\VT$ at large
ones. This regulates the infrared sensitivity
of the potential that has 
been widely discussed in the context of 
heavy-quark mass definitions.  

In order to combine the threshold behaviour from 
\eqs\nr{factor2}--\nr{expl} with the asymptotics
from \eq\nr{tilde_def_v}, we normalize $ \rho^\rmii{NRQCD}_\rmii{P} $
by a multiplicative factor. Given that the perturbative value of this
factor is poorly determined,  
we rather choose a free coefficient, denoted by $A$: 
\be
 \rho^\rmii{QCD}_\rmii{P} = 
   A \times 
 \rho^\rmii{NRQCD}_\rmii{P}
 \;. \la{A_def}
\ee
The value of $A$ is chosen so that 
$  \rho^\rmii{QCD}_\rmii{P} $
attaches to the $\msbar$ asymptotics from \eq\nr{tilde_def_v} 
continuously and with
a continuous first derivative at some 
$2 M < \omega < 3 M$, cf.\ \fig\ref{fig:inter}.

\begin{figure}[t]


\centerline{%
 \epsfysize=7.5cm\epsfbox{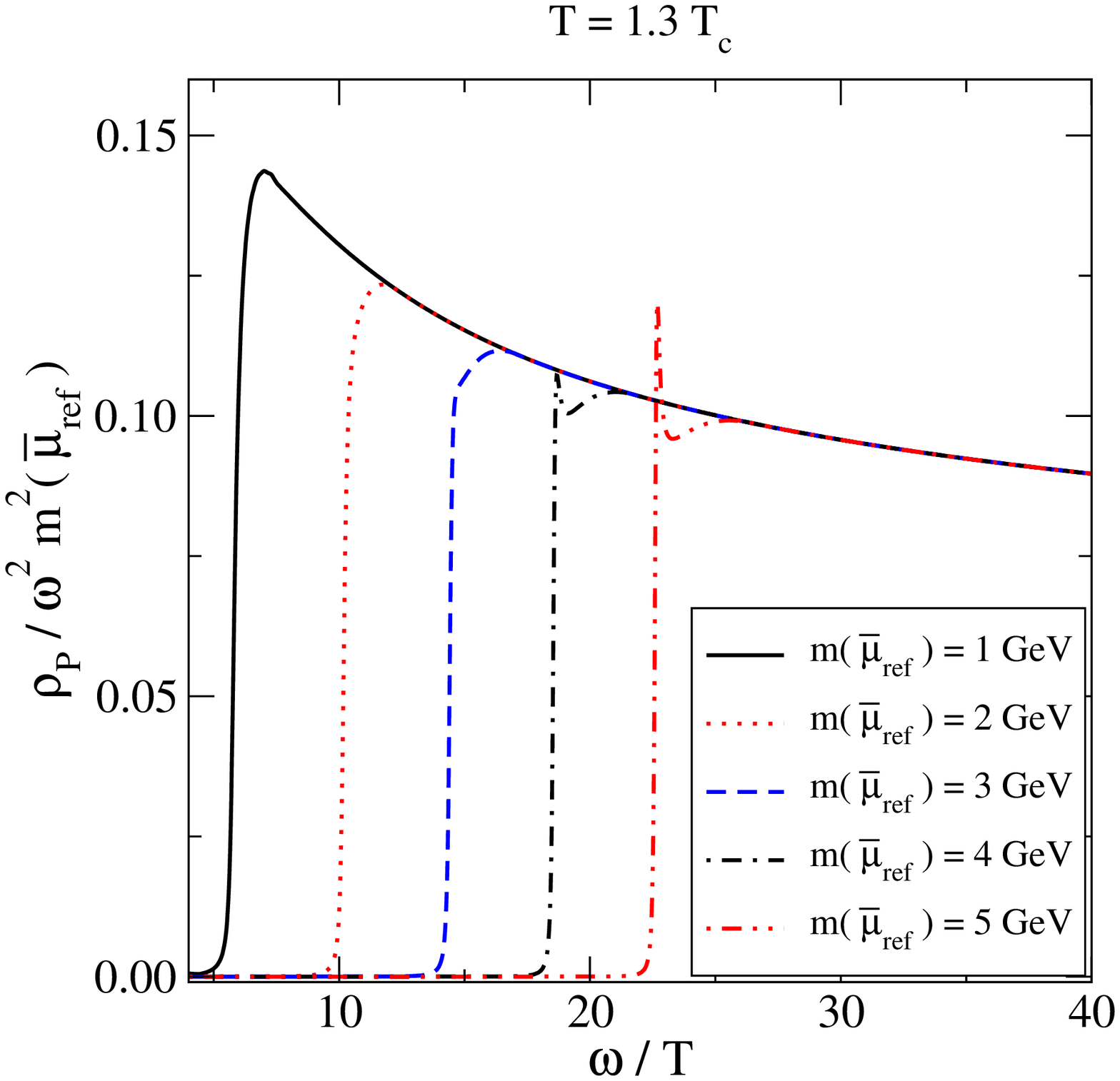}%
~~~\epsfysize=7.5cm\epsfbox{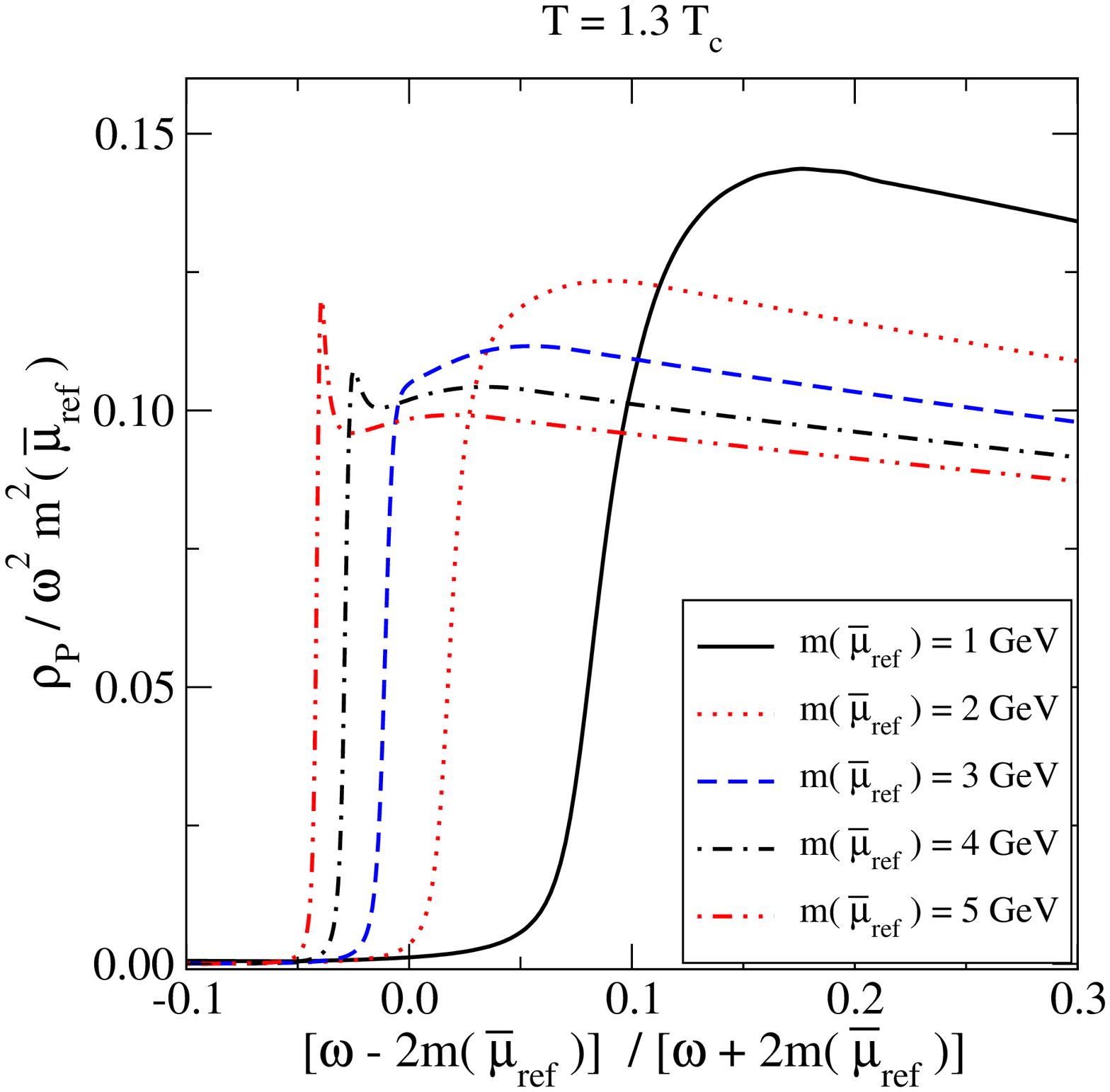}
}

\caption[a]{\small
 Spectral functions interpolating between a thermal contribution
 around the threshold (\se\ref{se:thermal}) and 
 $\msbar$ asymptotics above the threshold (\eq\nr{tilde_def_v}).
 The normalization of the near-threshold contribution 
 (cf.\ \eq\nr{A_def})
 and the matching 
 point have been treated as free parameters;  
 we find $A \approx 1.2 ... 1.3$ and 
 $\omega^{ }_\rmi{match} \approx 2.3 ... 2.6 M$. The other 
 parameters correspond to table~\ref{table:masses}, 
 with $M \equiv M^{ }_{x=6}$. 
}

\la{fig:inter}
\end{figure}

Once we are sufficiently {\em below} the threshold, 
$2M - \omega \gg \alphas^2 M$, 
the description of the spectral function through 
\eqs\nr{factor2}--\nr{expl} is no longer applicable. In fact the Schr\"odinger
description overestimates the spectral function in this regime: the 
correct result is non-zero at finite temperature, but exponentially
suppressed, as discussed below \eq\nr{full_TS}. In order to model
this suppression, we multiply $\phi$ of \eq\nr{expl} by
$\theta(2M - \omega) e^{-|\omega - 2M|/T}$. In the parametric range 
$|\omega - 2 M| \sim \alphas^2 M \ll gT, \alphas M, \pi T \ll M$ for which
an approach based on \eqs\nr{factor2}--\nr{expl} 
can be justified~\cite{jacopo,peskin,soto},
this amounts to a higher-order effect. 
The range $\omega \le 1.0 M$, 
which has very little effect after adopting this recipe, 
is cut off from the numerical evaluation of \eq\nr{relation}. 
These choices are purely phenomenological; formally they amount to 
effects which are of higher order in $\alphas$ than our computation. 

For small frequencies, 
the prefactor in \eq\nr{nrqcd_T} results in 
an additional suppression of the spectral function. However other
exponentially small (when $\omega \sim 2 M \gg \pi T$)
effects have been omitted in the non-relativistic 
solution of $C^{ }_{>}$. We ``reconstruct'' the exponential 
factor by taking it over from the tree-level 
computation in which, as indicated by \eq\nr{GPSfree}, it
is given by $\tanh(\frac{\omega}{4T}) = 1 - 2 \nF{}(\omega/2)$.
For $m(\bmu^{ }_\rmi{ref}) = 1$~GeV, 
when the non-relativistic
expansion is at most marginally viable, this leads
to a $\sim 10$\% reduction of the threshold enhancement
obtained from the Schr\"odinger equation.  
Of course, this reduction is partly compensated for by the
matching in \eq\nr{A_def}. 
In any case, our ``best estimates'' for 
the thermal spectral function 
at different $m(\bmu^{ }_\rmi{ref})$, obtained as outlined above,  
are shown in \fig\ref{fig:inter}. 

 We note in passing that, in accordance with the original findings of 
 ref.~\cite{peskin}, a small resonance peak can be observed
 at the largest quark
 masses at this temperature. We return to a discussion of this point 
 in \se\ref{se:model}.

%
\section{Lattice simulations}
\la{se:lattice}

%
\subsection{Ensemble and statistics}

We have carried out lattice simulations in quenched SU(3) gauge 
theory with $\rmO(a)$ improved Wilson quarks as valence quarks, 
in order to measure 
the connected pseudoscalar correlator and take the continuum limit.
At each lattice spacing, this involves an interpolation 
to a bare quark mass that reproduces the physical $J/\psi$ or $\Upsilon$
mass at a low temperature.  
The basic techniques (action, simulation algorithm), as well as
tests suggesting that finite-volume effects are reasonably small, are 
described in ref.~\cite{ding2}. The progress in the past
5 years amounts to simulating several 
fine lattice spacings, and on each of them 
several values of the bare quark mass. 
Details concerning the ensemble are collected
in table~\ref{table:params}.

%
\begin{table}[tp]

\small{
\begin{center}
\begin{tabular}{cccccccccc} 
 $\beta$ & 
 $\Ns$ &  $\Nt$ &
 confs & 
 $ \rO / a $ & 
 $ T / \Tc  $ & $c^{ }_\rmii{SW}$ & $\kappa^{ }_\rmi{c}$ & 
 $ \kappa $ &
  \hspace*{-0.2cm} $\frac{m^2({1}/{a})}{m^2(\bmu^{ }_\rmi{ref})}$ 
 \\[3mm]
 \hline 
  7.192  & 96 &  48 & 237 
    & 26.6 & 0.74 & 1.367261 & 0.13442 &     
    {\tiny  
    $\begin{array}{l}
     0.12257,0.12800,0.13000, \\ 0.13100,0.13150,0.13194
    \end{array}$
    } & 0.6442 \\[-1mm] 
         &     &  32 & 476     &    & 1.12 & & & & \\ 
         &     &  28 & 336     &    & 1.27 & & & & \\ 
         &     &  24 & 336     &    & 1.49 & & & & \\ 
         &     &  16 & 237     &    & 2.23 & & & & \\ 
  7.394  & 120 & 60 & 171  
    & 33.8 & 0.76 &  1.345109 & 0.13408 & 
    {\tiny  
    $\begin{array}{l}
     0.124772,0.12900,0.13100, \\ 0.13150,0.132008,0.132245
    \end{array}$
    }
    &  0.6172 \\[-1mm] 
         &     &  40 & 141     &    & 1.13 & & & & \\ 
         &     &  30 & 247     &    & 1.51 & & & & \\ 
         &     &  20 & 226     &    & 2.27 & & & & \\ 
  7.544  & 144 & 72 & 221 
    & 40.4 & 0.75 & 1.330868 & 0.13384 & 
    {\tiny
     $\begin{array}{l}
      0.12641,0.12950,0.13100, \\ 0.13180,0.13220,0.13236
    \end{array}$
     } &  0.5988 \\[-1mm] 
         &     &  48 & 462     &    & 1.13 & & & & \\ 
         &     &  42 & 660     &    & 1.29 & & & & \\ 
         &     &  36 & 288     &    & 1.51 & & & & \\ 
         &     &  24 & 237     &    & 2.26 & & & & \\ 
  7.793  & 192 & 96 & 224 
    & 54.1 & 0.76 & 1.310381 & 0.13347 & 
    {\tiny
       $\begin{array}{l}
      0.12798,0.13019,0.13125, \\ 0.13181,0.13209,0.13221
     \end{array}$
    }  & 0.5715 \\[-1mm] 
         &     &  64 & 249     &    & 1.13 & & & & \\ 
         &     &  56 & 190     &    & 1.30 & & & & \\ 
         &     &  48 & 210     &    & 1.51 & & & & \\ 
         &     &  32 & 235     &    & 2.27 & & & & \\ 
 \hline 
\end{tabular} 
\end{center}
}

\vspace*{3mm}

\caption[a]{\small
  The lattices (of geometry $\Ns^3\times\Nt$)
  included in the current analysis. 
  The measurements of $\rO/a$ at low temperatures are based on 
  our own measurements of large Wilson loop expectation values, 
  and correspond to a sector of fixed trivial topology, due to 
  the freezing of topological degrees of freedom~\cite{slow}; 
  we refrain from estimating systematic uncertainties.
  When citing physical units, 
  we make use of $\rO = 0.47(1)$~fm from ref.~\cite{r0fm}. 
  Conversions to units of $\Tc$ are based on 
  $\rO \Tc = 0.7457(45)$ from ref.~\cite{betac}.
  Our new data for $\rO/a$, together with the older one summarized
  in table~2 of ref.~\cite{betac},
  which include classic results from refs.~\cite{r0old1,r0old2}, 
  can be well represented 
  ($\chi^2\mbox{/d.o.f.} = 1.43$) 
  by \eq(4) of ref.~\cite{betac}, with the modified couplings 
  $
    c^{ }_1 = -8.9664
  $, 
  $ 
    c^{ }_2 = 19.21
  $, 
  $
    c^{ }_3 = -5.25217 
  $, 
  $
    c^{ }_4 = 0.606828
  $.  
  The Sheikholeslami-Wohlert coefficient $c^{ }_\rmii{SW}$~\cite{cSW}
  and the critical Wilson hopping parameter $\kappa^{ }_\rmi{c}$ are 
  in accordance with ref.~\cite{Oa}; the bare lattice mass is 
  $a m^{ }_\rmii{L} = \frac{1}{2\kappa} - \frac{1}{2\kappa^{ }_\rmi{c}}$. 
  The evolution of the $\msbar$ renormalized quark mass squared
  given in the last column, 
  which affects the renormalized pseudoscalar correlator as described in 
  \se\ref{ss:renorm}, 
  was determined with 5-loop running~\cite{mass1,mass2,alph1,alph2}, 
  with $\bmu^{ }_\rmi{ref} \equiv 2$~GeV.   
 }
\label{table:params}
\end{table}
%

There is a well-known problem with simulations close to the 
continuum limit with periodic boundary conditions, namely the
``freezing'' of topological degrees of freedom~\cite{slow}. 
At low temperatures, this effect is unphysical. 
Our measurements were separated by 500 sweeps, each consisting
of 1 heatbath and 4 overrelaxation updates. The initial thermalization
consisted of 2000 sweeps at $\beta = 7.192$ and of 5000 sweeps
at $\beta = 7.793$. With this setup we
do observe some evolution between the topological sectors up to
$\beta \simeq 6.8$, however at the values $\beta \gsim 7.2$ that
play a role in our analysis, no evolution takes place. Even though
the resulting uncertainties should be minor at high temperatures, 
where the physical topological susceptibility is small, the 
freezing does affect our low-temperature runs as well, 
particularly our scale setting~\cite{betac}\footnote{%
   We take the opportunity to note that the values of 
   $(t^{ }_0/a^2)^\rmii{Clover}$ cited in ref.~\cite{betac} were
   marred by a bug, and are too large by a few \%. This does
   not affect any of the conclusions of ref.~\cite{betac}, because
   the clover results were excluded due to their peculiar
   volume dependence. We thank Lukas Mazur for locating the bug.  
  }
though the parameter $\rO$~\cite{r0}.
In some sense, the values of $\rO/a$ in our study 
should be interpreted as ``$\rO/a$ at fixed trivial topology''.  
It is not known how much these differ from the 
corresponding $\rO/a$ in the physical $\theta$-vacuum, however
in principle the differences are suppressed by the inverse volume and
therefore perhaps moderate. 

%
\subsection{Tuning of quark mass}

\begin{figure}[t]


\centerline{%
 \epsfysize=7.3cm\epsfbox{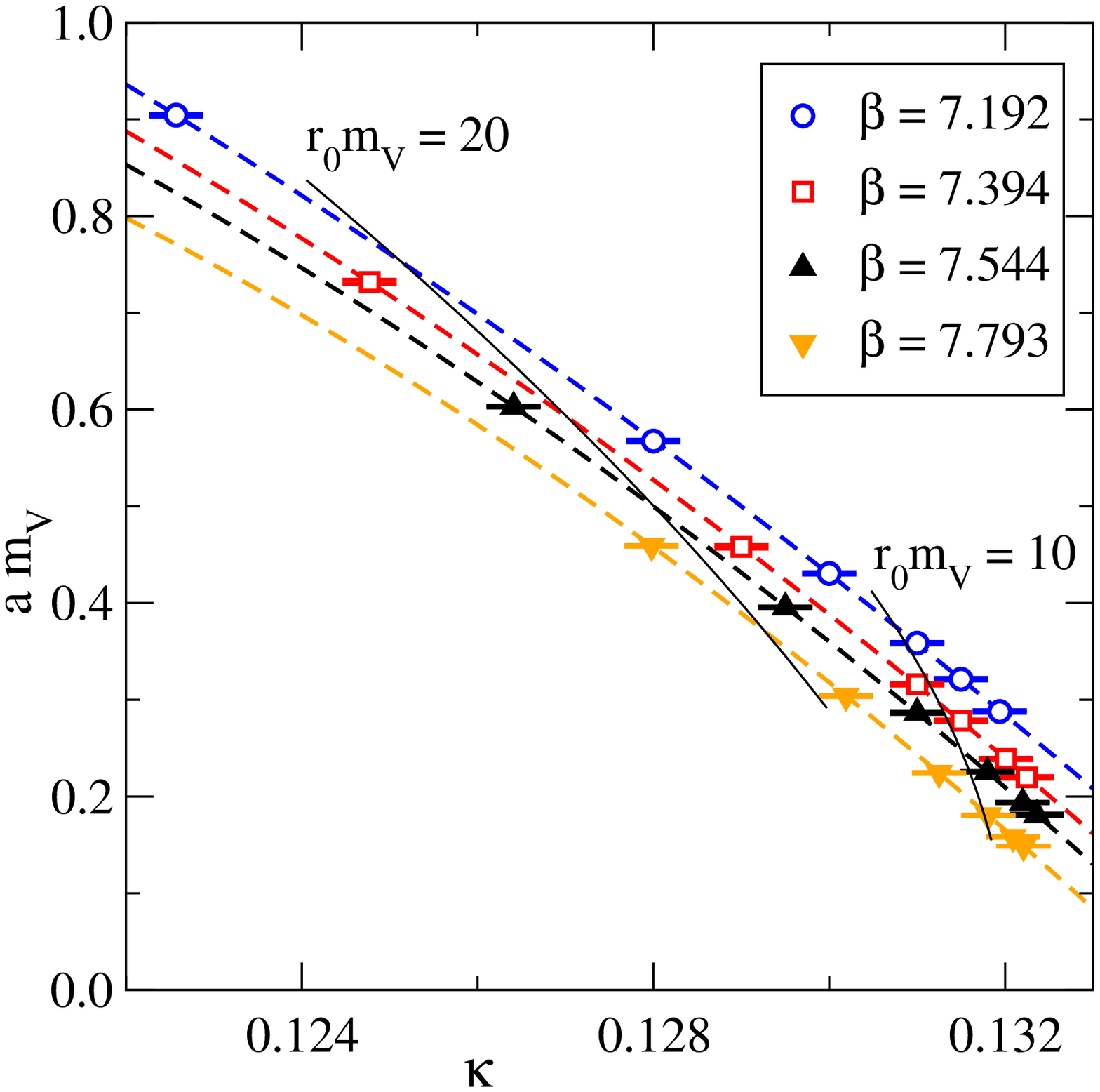}%
~~~\epsfysize=7.9cm\epsfbox{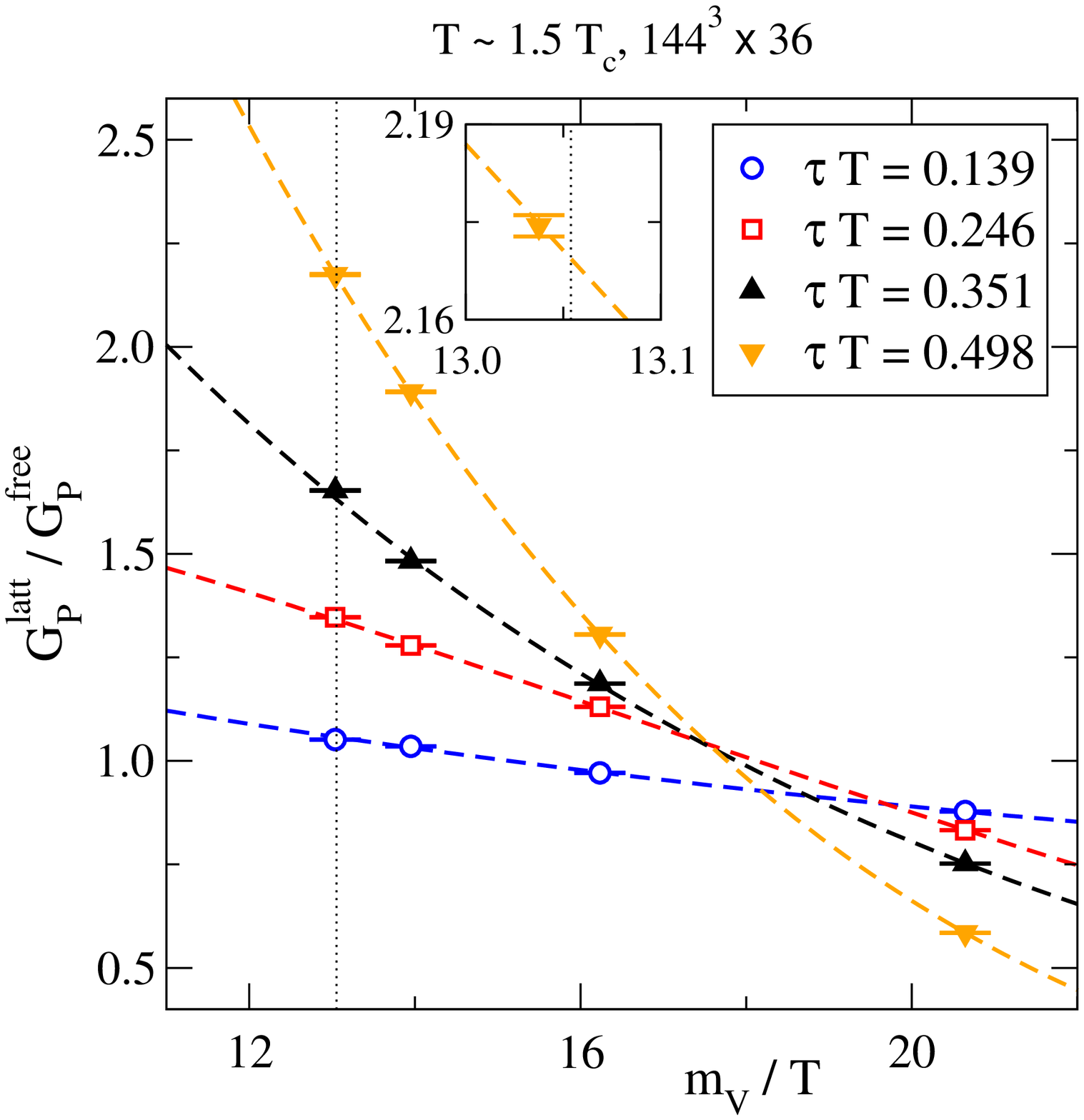}
}

\caption[a]{\small
 Left: The vector meson (physically $J/\psi,\Upsilon$) masses
 in lattice units, $a m_\rmii{V}^{ }$,  measured from spatial 
 correlators deep in the confined phase. Dashed lines 
 show quadratic fits, meant to guide the eye. 
 Solid lines show ``lines of constant physics'', with the 
 conversion of $a$ to physical units being based on $\rO$, 
 cf.\ table~\ref{table:params}
 ($\rO m^{ }_\rmii{V} = 20$ corresponds to $m^{ }_\rmii{V} \simeq$~8~GeV,
  $\rO m^{ }_\rmii{V} = 10$ to $m^{ }_\rmii{V} \simeq$~4~GeV).    
 Right: An example of an interpolation of the thermal correlator
 to a physical mass value (in this case 
 $m^{ }_\rmii{$J/\psi$}\approx 3.1$~GeV, 
 indicated by a dotted line). 
 Units are again converted via $\rO\approx 0.47$~fm~\cite{r0fm}.
}

\la{fig:masses}
\end{figure}

For any given lattice spacing, the first step is to interpolate
results to the physical bare quark mass. We have determined quark
propagators at six
different bare quark masses (or hopping parameters; 
cf.\ table~\ref{table:params}), roughly speaking between the 
charm and bottom masses, with a denser grid around
the former. As the corresponding observable
we measure the screening mass corresponding to the spatial components
of the vector current (including only transverse components with
respect to the measurement direction).  
The measurement is carried out deep in the confined phase
($T \sim 0.75\Tc$ in table~\ref{table:params}), and is 
interpreted as being a good approximation 
for a measurement at zero temperature. 
Thus, the screening mass is identified as the $J/\psi$ or 
$\Upsilon$ mass (we use $J/\psi$ rather than $\eta^{ }_c$ for
scale setting because it is a narrower resonance, however in 
practice this difference is inessential on our resolution). 
The screening mass is extracted from a two-exponential fit,
with an ensemble of fit ranges chosen by hand close to where 
a minimal $\chi^2/\mbox{d.o.f.}\sim 1$ can be found. 
The result is denoted by $m^{ }_\rmii{V}$; the corresponding values 
are plotted in \fig\ref{fig:masses}(left) in lattice units. 

Proceeding now to the thermal runs in the temporal direction ($\tau$), 
the correlators were measured at the same values
of the bare quark masses. Therefore each measurement can be assigned
to a specific value of the zero temperature
(screening) mass $m^{ }_\rmii{V}$. 
The data is subsequently interpolated quadratically in~$m^{ }_\rmii{V}$, 
\be
 \ln \biggl\{ 
   \frac{ G^\rmi{latt}_\rmii{P}(\tau T) }{ G^\rmi{free}_\rmii{P}(\tau T) }
 \biggr\} 
 = 
 \alpha^{ }_2(\tau T)\, \biggl(\frac{m^{ }_\rmii{V}}{T}\biggr)^2 + 
 \alpha^{ }_1(\tau T)\, \frac{m^{ }_\rmii{V}}{T} + 
 \alpha^{ }_0(\tau T)
 \;, \la{interpolation}
\ee
where $G^\rmi{free}_\rmii{P}$ is defined in \eq\nr{GPSfree}. 
The four $\kappa$-values closest to the desired point are used
for this interpolation (or, in some cases, mild extrapolation). 
Having determined the coefficients $\alpha^{ }_i(\tau T)$, 
we finally set $m^{ }_\rmii{V} \to m^{ }_\rmii{$J/\psi$}$
or $m^{ }_\rmii{V} \to m^{ }_\rmii{$\Upsilon$}$, 
in order to get the physical result at the given $\tau$ and $T$. 
The procedure is illustrated in \fig\ref{fig:masses}(right) for 
four values of $\tau T$. The interpolated correlator for
the physical mass can be read off from the dotted line.  

%
\subsection{Normalization of imaginary-time correlators}

With the data interpolated to a physical quark mass, the next task 
is to extrapolate to the continuum limit. This is facilitated 
by normalizing the lattice data to a function which captures most of the 
rapid $\tau$-dependence at small $\tau \ll 1/T$. 

Because of a non-zero anomalous dimension, 
there is no exact ``free result'' 
for the pseudoscalar correlator like, say, 
for the vector channel correlator in the chiral limit. 
However, we can {\em define} 
\ba
 \frac{ G_\rmii{P}^\rmi{free}(\tau) }{ m^2(\bmu^{ }_\rmi{ref}) } 
 & \equiv &   
 \int_{2 M^{ }_\rmii{1S}}^{\infty}
 \! \frac{{\rm d}\omega}{\pi} \, 
 \biggl\{ 
 \frac{\Nc \omega^2}{8\pi}
 \tanh\Bigl( \frac{\omega}{4T} \Bigr)  
 \sqrt{1 - \frac{4 M_\rmii{1S}^2}{\omega^2}} 
 \biggr\}
 \, 
 \frac{\cosh \left(\left(\frac{1}{2T} - \tau\right)\omega\right)}
 {\sinh\left(\frac{\omega}{2T}\right)} 
 \;, \la{GPSfree}
\ea
where
the expression in curly brackets is the tree-level spectral
function, and the remaining factors amount to those in \eq\nr{relation}. 
Because of the anomalous dimension, 
a result normalized through \eq\nr{GPSfree} does not go 
to a constant value at $\tau \ll 1/T$, 
though we expect it to be slowly varying. 
By definition we set $M^{ }_\rmii{1S} \equiv 1.5$~GeV
in \eq\nr{GPSfree} for the charmonium case, 
and $M^{ }_\rmii{1S} \equiv 5.0$~GeV for the bottomonium case
(the physical value being $M^{ }_\rmii{1S} \approx 4.7$~GeV).

%
\subsection{Renormalization and continuum extrapolation}
\la{ss:renorm}

In the absence of non-perturbative renormalization factors
for quenched massive pseudoscalar densities, we have made use of 
massless perturbative results~\cite{sc,latZ}, 
supplemented by a 1-loop correction for the mass dependence~\cite{aM0,aM}. 
A tadpole-improved coupling was inserted in these formulae,
along the lines discussed in ref.~\cite{go1}.
The difference of the 1-loop and 2-loop expressions from 
ref.~\cite{latZ} was employed for getting a feeling about 
the systematic uncertainties of perturbative renormalization. 

There is an important subtlety related to the 
pseudoscalar renormalization factor, 
denoted by $Z^{ }_\rmii{P}$. 
Since $Z^{ }_\rmii{P}$ is supposed to bring us from
the lattice to the $\msbar$ scheme, it depends on how $\gamma^{ }_5$ 
is defined on the latter side. With the 't Hooft-Veltman
choice~\cite{hv}, the strict $\msbar$ 
pseudoscalar density needs to be multiplied with 
an additional finite renormalization factor in order for it to have 
the same anomalous dimension as the scalar density~\cite{larin}, 
\be
 Z^{ }_5 =
 1 - \frac{g^2 \CF^{ }}{2\pi^2} + \frac{g^4 \CF^{ }}{128\pi^4}
 \frac{\Nc + 2 \Nf}{9}
 + \rmO(g^6)
 \;. \la{larin}
\ee
In so-called naive dimensional regularization, no additional factor 
is needed at low perturbative orders. 
The 1-loop pseudoscalar renormalization factor computed in 
ref.~\cite{sc} is smaller than that in ref.~\cite{latZ}, 
the reason being that it is multiplied by $Z^{ }_5$
in comparison with ref.~\cite{latZ}. 
We are interested
in a pseudoscalar density which 
corresponds 
to the perturbative result presented in \se\ref{se:vacuum}
and has the same
anomalous dimension as the scalar density (cf.\ comments
below \eq\nr{tR2l}).
Then the normalization of ref.~\cite{sc} is appropriate,
i.e.\ we need $Z^{ }_5 Z^\rmii{L,$\msbar$}_\rmii{P}$ in 
the notation of ref.~\cite{latZ}. For reference we note
that numerically $Z^{ }_5 Z^\rmii{L,$\msbar$}_\rmii{P} \sim 0.8$
in the range of our $\beta$'s and masses.

The perturbative renormalization factors bring us
to the $\msbar$ scheme 
at the scale $\bmu = 1/a$. For scale-independent results
the pseudoscalar correlator should then
be multiplied by a quark mass in the same scheme and
at the same renormalization scale, i.e.\ $m^2(1/a)$. 
Subsequently the mass can 
be evolved to the scale $\bmu^{ }_\rmi{ref}$ so that it  
drops out if the correlator is normalized according to 
\eq\nr{GPSfree}: 
$m^2(a^{-1}) = m^2(\bmu^{ }_\rmi{ref})\, [m(a^{-1})/m(\bmu^{ }_\rmi{ref})]^2$. 
In other words, lattice results multiplied by 
$Z^{ }_5 Z^\rmii{L,$\msbar$}_\rmii{P}$ need to be further multiplied
by $[m(a^{-1})/m(\bmu^{ }_\rmi{ref})]^2$;   
the numerical value of this ratio is presented in table~\ref{table:params}.

\begin{figure*}


\centerline{%
 \epsfysize=7.5cm\epsfbox{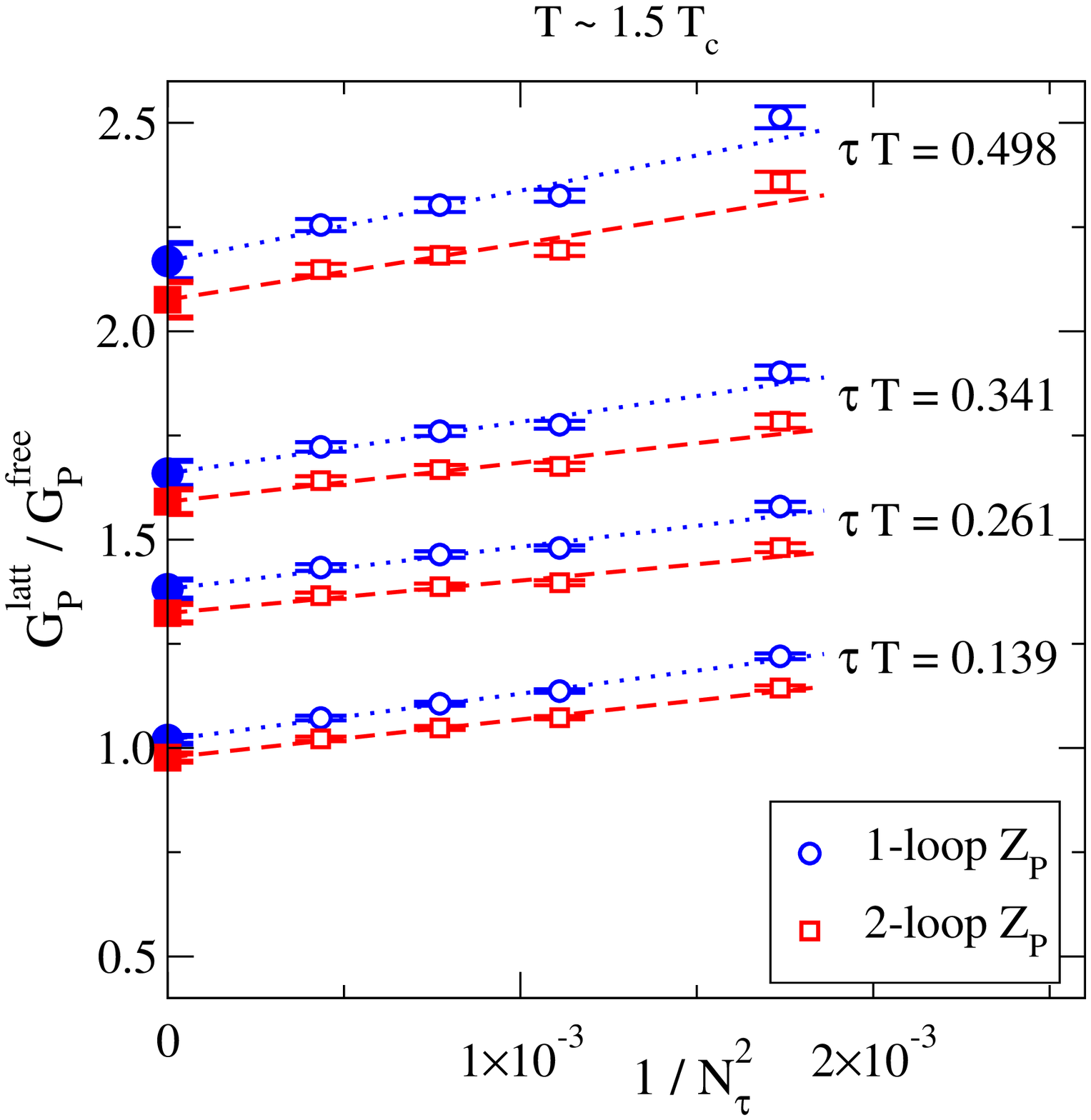}
~~~\epsfysize=7.5cm\epsfbox{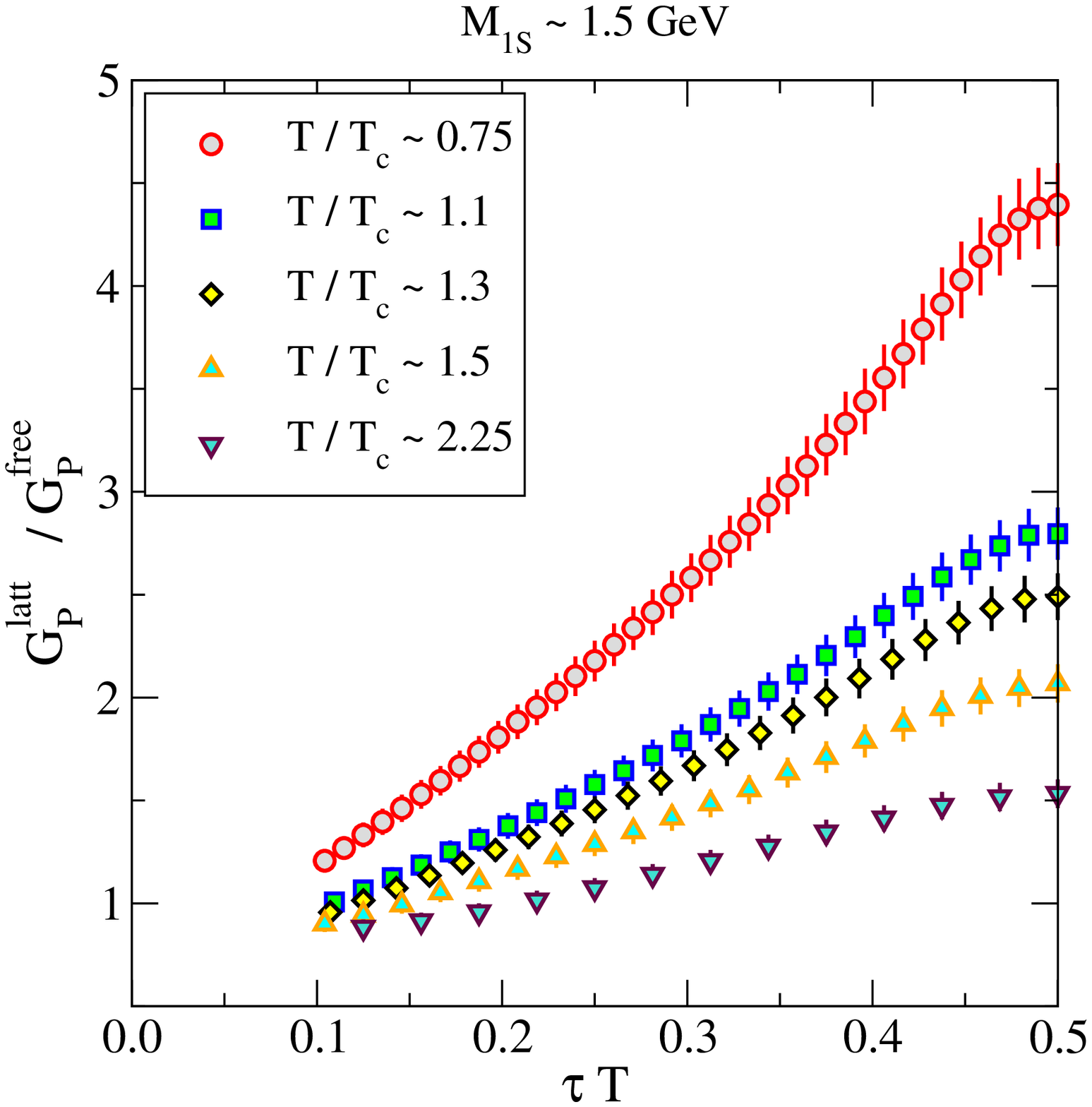}%
}

\caption[a]{\small
Left: Examples of continuum extrapolations at $T\sim1.5\Tc$
(filled symbols), 
based on 1-loop and 2-loop renormalization factors and a fit
ansatz linear in $1/N_\tau^2$ (cf.\ \se\ref{ss:renorm}). We have also 
carried out fits quadratic in $1/N_\tau^2$, or linear in $1/N_\tau^2$
but restricted to the three finest lattices; the variations are of 
the same order as the differences from using 1-loop and 2-loop 
renormalization factors, which we then adopt as our estimate
of systematic uncertainties. 
Right: Continuum-extrapolated correlators. The uncertainties 
have been obtained by adding together, in quadrature, 
statistical errors and systematic uncertainties
from the difference of using 1-loop and 2-loop renormalization factors.
} 

\la{fig:taudep}
\end{figure*}

\begin{figure*}


\centerline{%
 \epsfysize=7.5cm\epsfbox{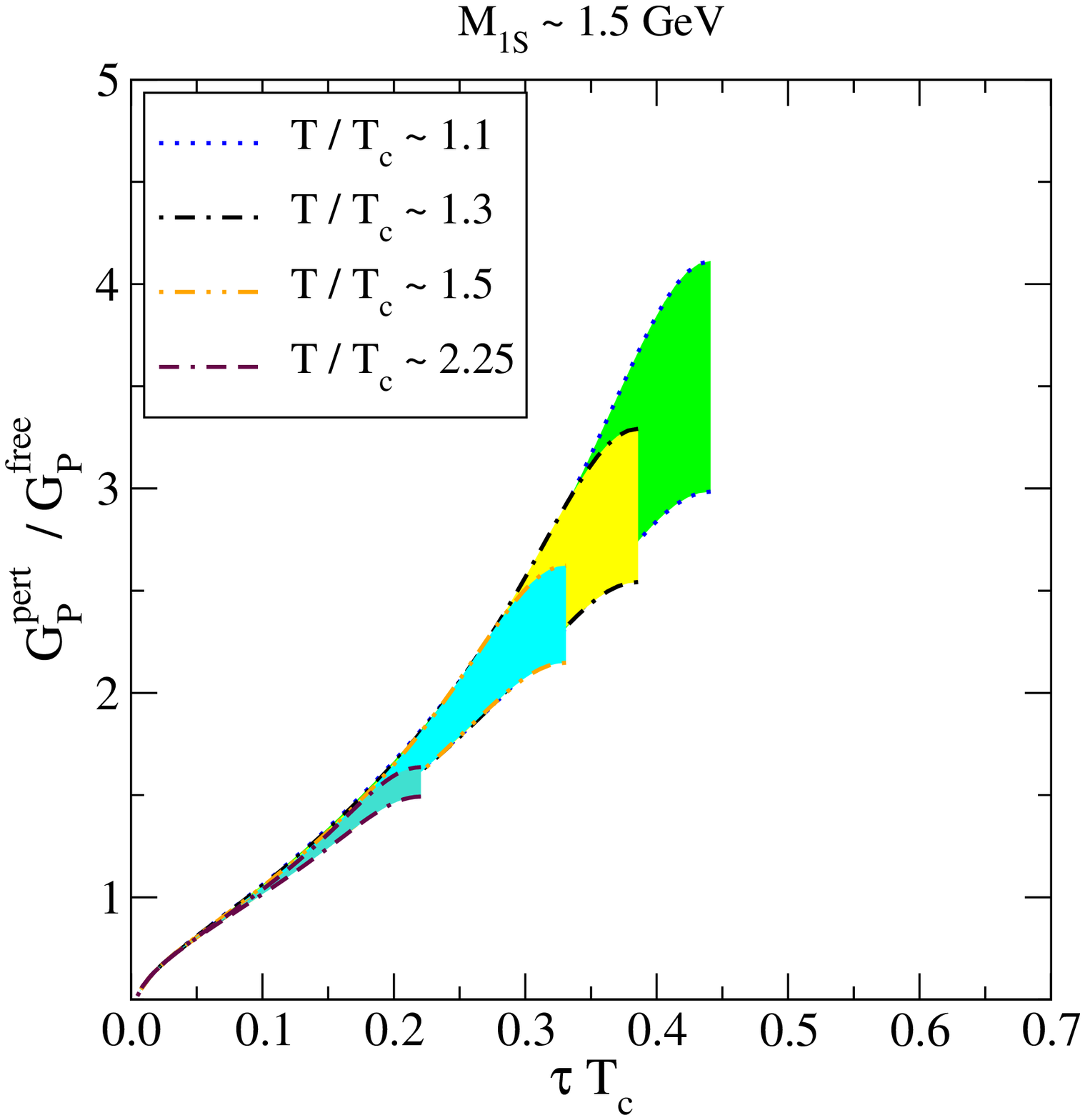}
~~~\epsfysize=7.5cm\epsfbox{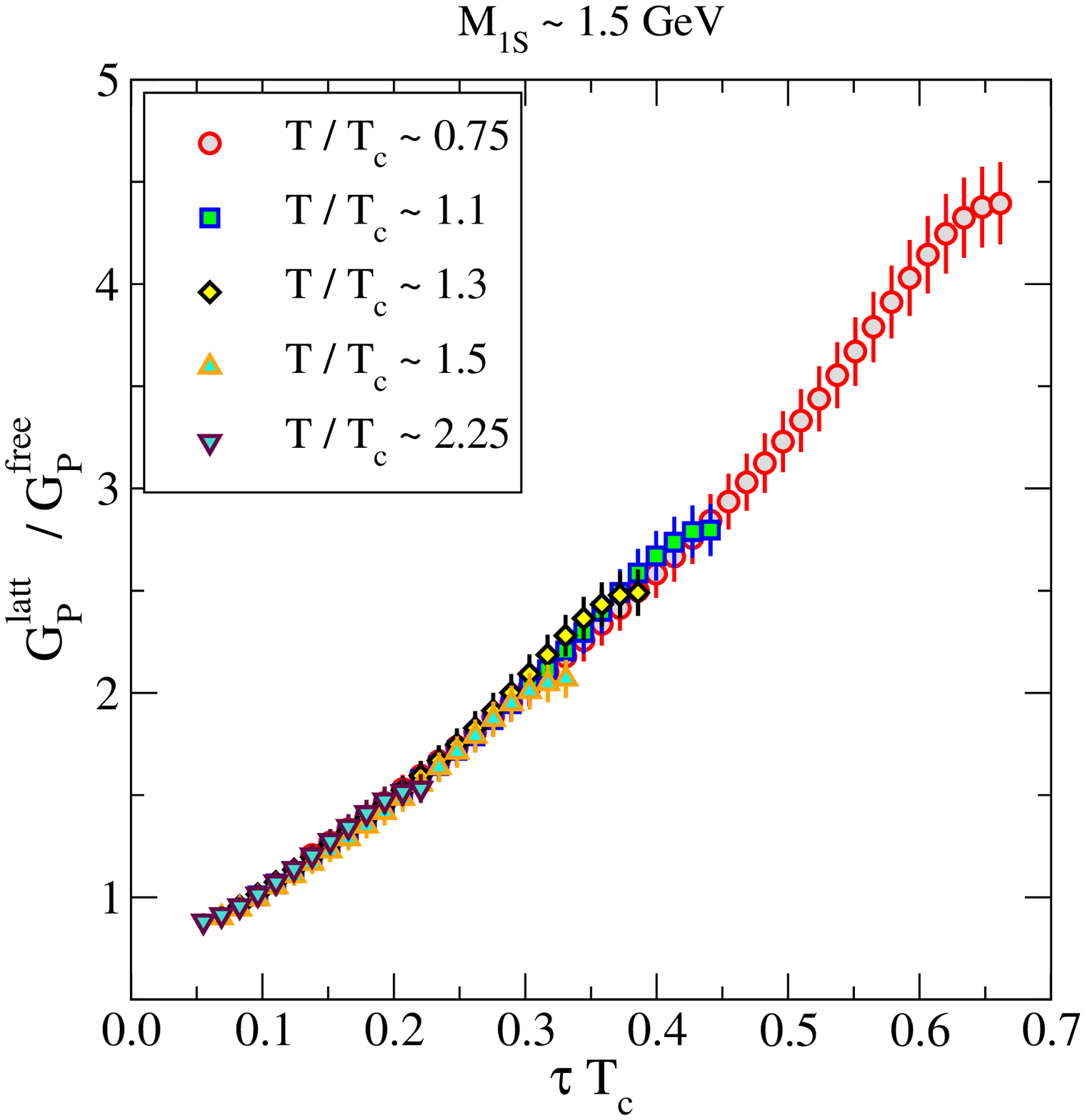}%
}

\caption[a]{\small
Left: Imaginary-time correlators corresponding 
to spectral functions such as those shown in \fig\ref{fig:inter}, 
for $m(\bmu^{ }_\rmi{ref}) = 1$~GeV, 
plotted by expressing $\tau$
in fixed units, chosen as $\Tc$.
The uncertainty bands have been 
obtained by varying the central value of $m(\bmu^{ }_\rmi{ref})$ 
by 10\% in both directions.
Right: Lattice
data from fig.~\ref{fig:taudep}(right) in the same units, including
now also the lowest temperature at $T < \Tc$.   
}

\la{fig:taudep2}
\end{figure*}

\begin{figure*}


\centerline{%
 \epsfysize=7.5cm\epsfbox{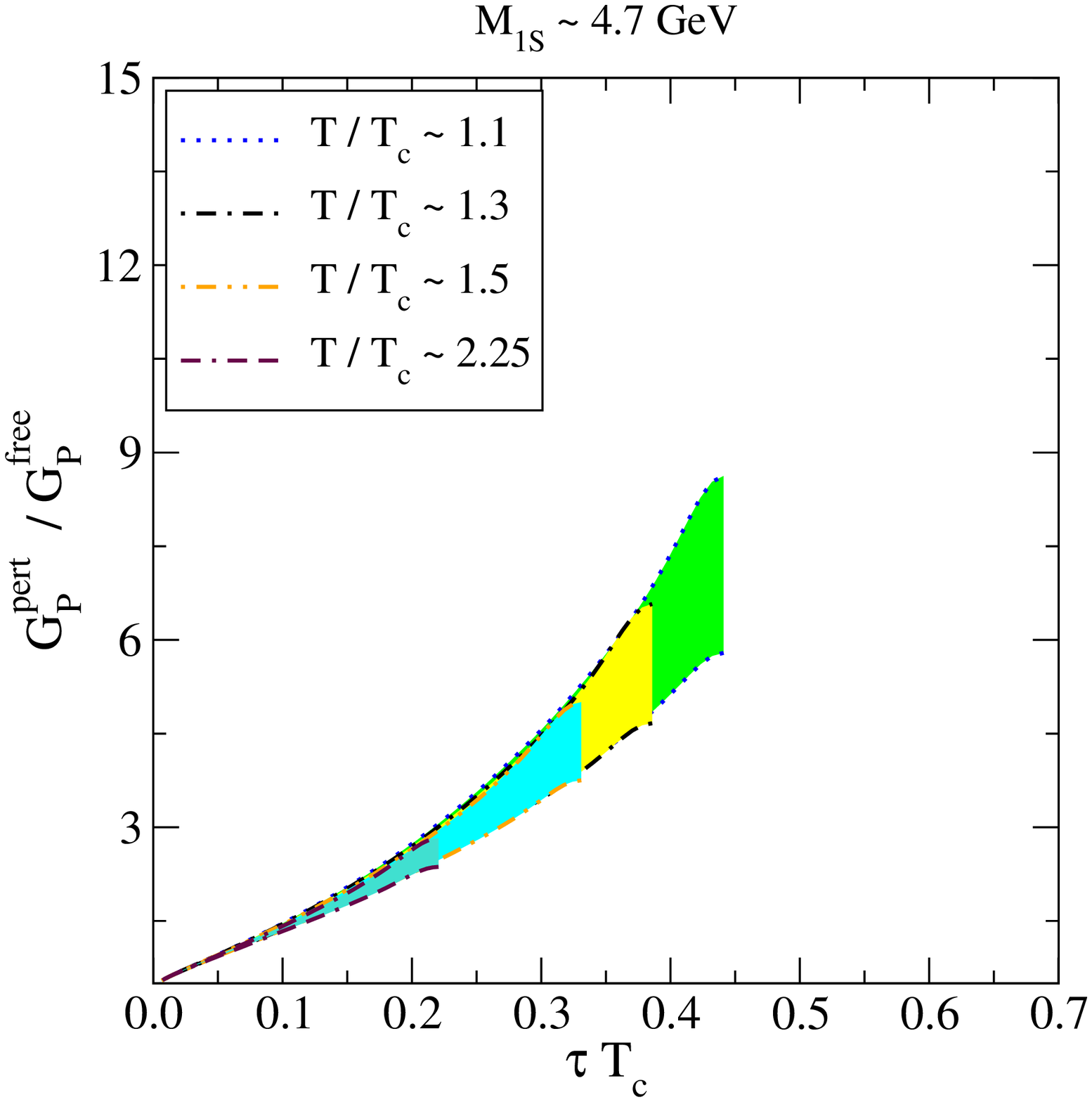}
~~~\epsfysize=7.5cm\epsfbox{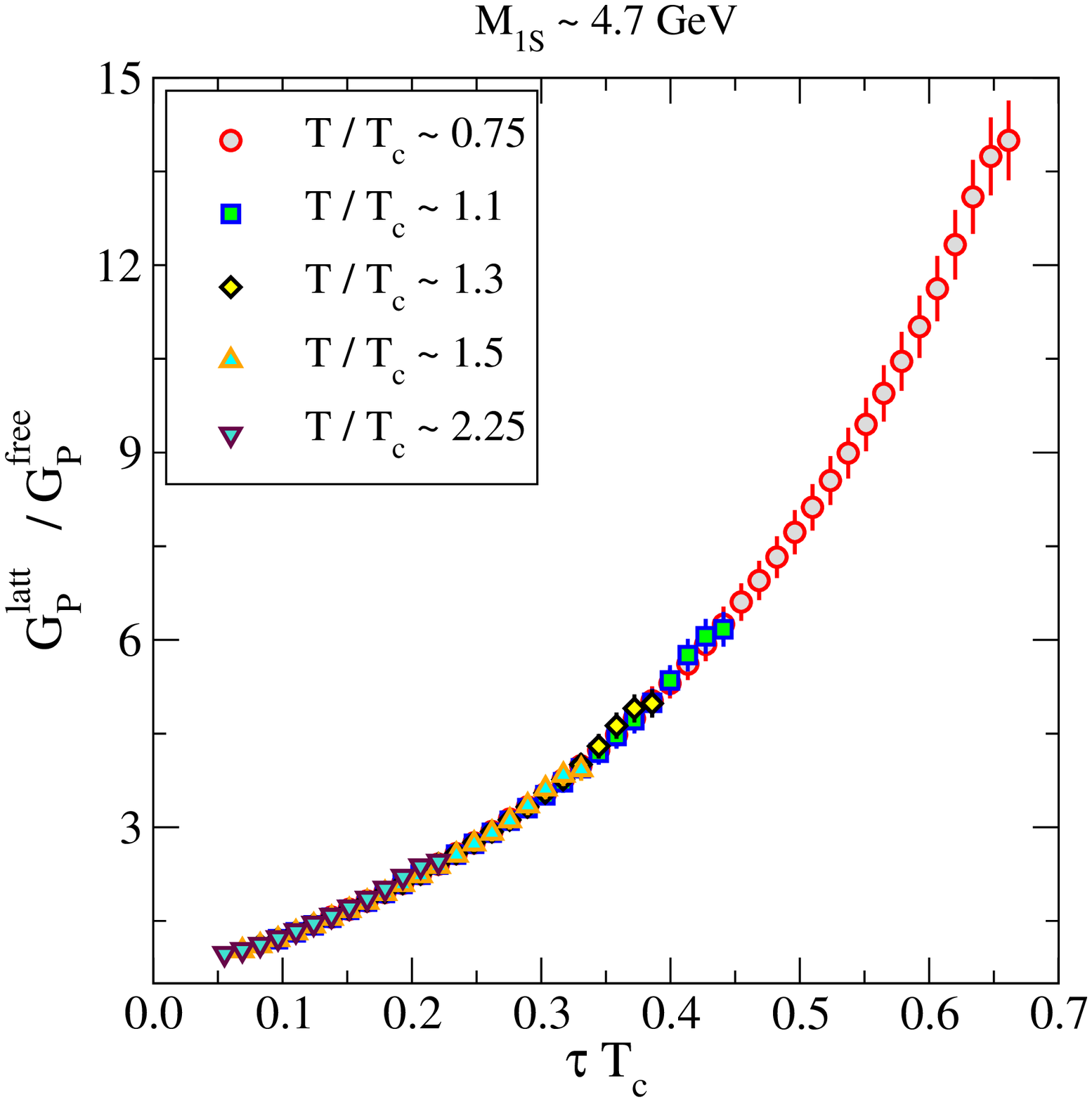}%
}

\caption[a]{\small
Like \fig\ref{fig:taudep2} but for the bottomonium case. 
Here the perturbative uncertainty bands have been 
obtained by varying the central value $m(\bmu^{ }_\rmi{ref}) = 5$~GeV 
by 2\% in both directions.
}

\la{fig:taudep2_bot}
\end{figure*}

After multiplication by the renormalization factors, 
piecewise polynomial interpolations (b-splines) 
are used for estimating all correlators 
at the distances available on the finest lattice. 
The continuum extrapolation is then carried out
independently at all these distances.
An ansatz linear in $1/N_\tau^2$ describes the data well, 
even though effects of $\rmO(\alphas^2 a m^{ }_\rmii{L})$ 
are thereby omitted~\cite{aM}
(here $a m^{ }_\rmii{L}$ is the bare lattice 
mass defined in the caption of table~\ref{table:params}).  
Representative examples of continuum extrapolations are 
shown in \fig\ref{fig:taudep}(left). All four lattice 
spacings, as listed in table~\ref{table:params},
are included in the fit, except for the temperature
$T \sim 1.3\Tc$ at which only three lattice spacings are available. 
Other fit forms (quadratic in $1/N_\tau^2$, or linear in $1/N_\tau^2$
but restricted to the three finest lattices) have also 
been successfully tested, cf.\ caption
of \fig\ref{fig:taudep}.

The whole procedure is implemented in the form
of a bootstrap analysis. The final continuum-extrapolated results
are shown in the right panels of \figs\ref{fig:taudep} and \ref{fig:taudep2}
for charmonium, and in the right panel of \fig\ref{fig:taudep2_bot} 
for bottomonium. 

%
\section{Modelling and comparison}
\la{se:model}

The purpose of this section is to compare the predictions
originating from the spectral functions in \fig\ref{fig:inter}, 
shown as the corresponding imaginary-time correlators for the 
charmonium case in \fig\ref{fig:taudep2}(left) and for
the bottomonium case in \fig\ref{fig:taudep2_bot}(left),  
with the lattice data shown in \figs\ref{fig:taudep2}(right) and 
\ref{fig:taudep2_bot}(right). We restrict ourselves to 
a rather qualitative discussion here, 
identifying two main issues which  
help to explain the differences observed. 

Let us start by noting that 
one striking feature of the lattice data is its
apparent inertness: if plotted in units of $\tau\Tc$, 
all curves more or less fall on top of each other, 
cf.\ \fig\ref{fig:taudep2}(right). 
Such an inertness, 
however,  does {\em not} indicate the 
persistence of bound states; indeed a similar inertness
is seen on the perturbative side
where no bound states are present, 
cf.\ \fig\ref{fig:taudep2}(left).

\begin{figure*}


\centerline{%
 \epsfysize=7.5cm\epsfbox{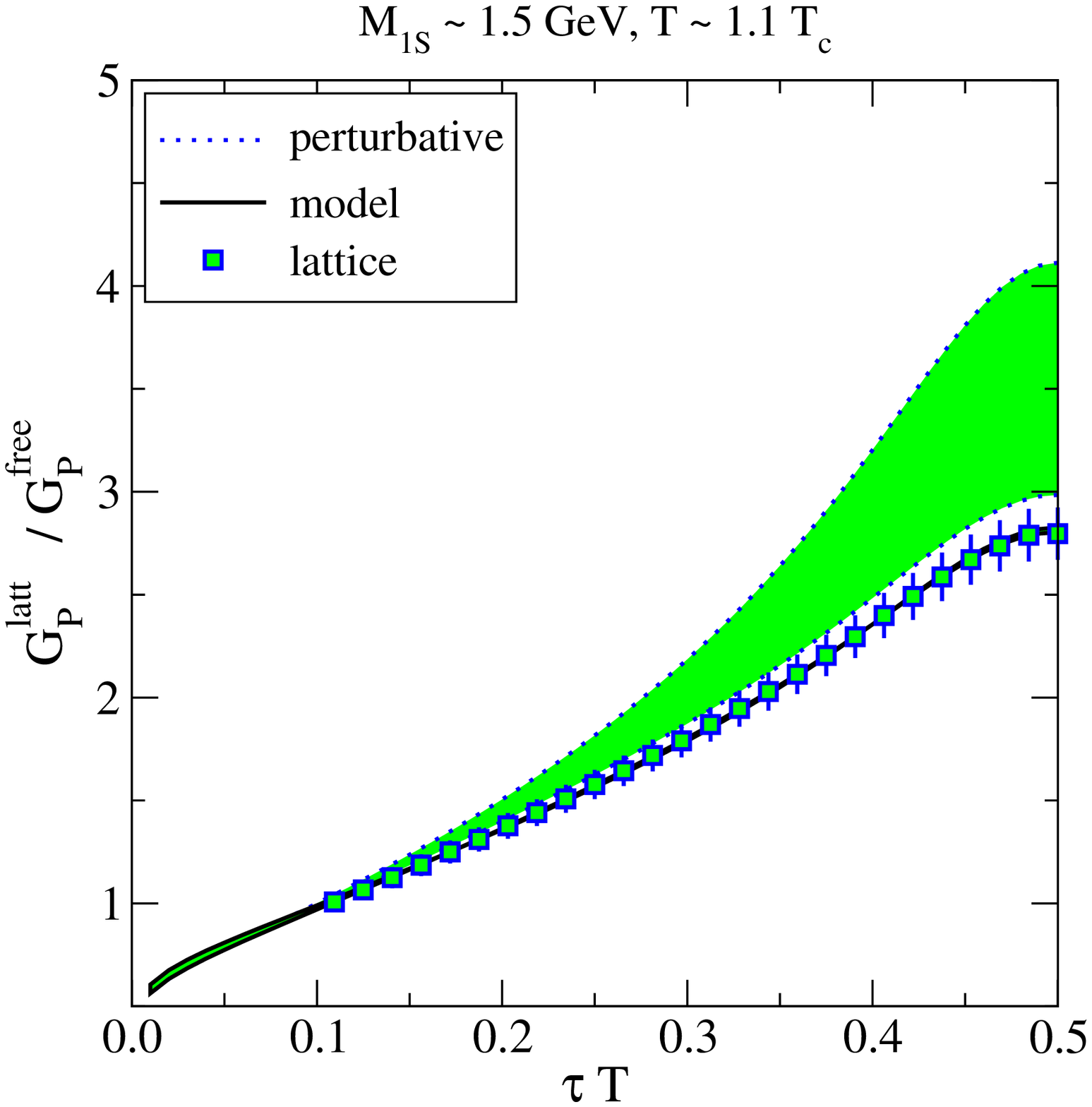}
~~~\epsfysize=7.5cm\epsfbox{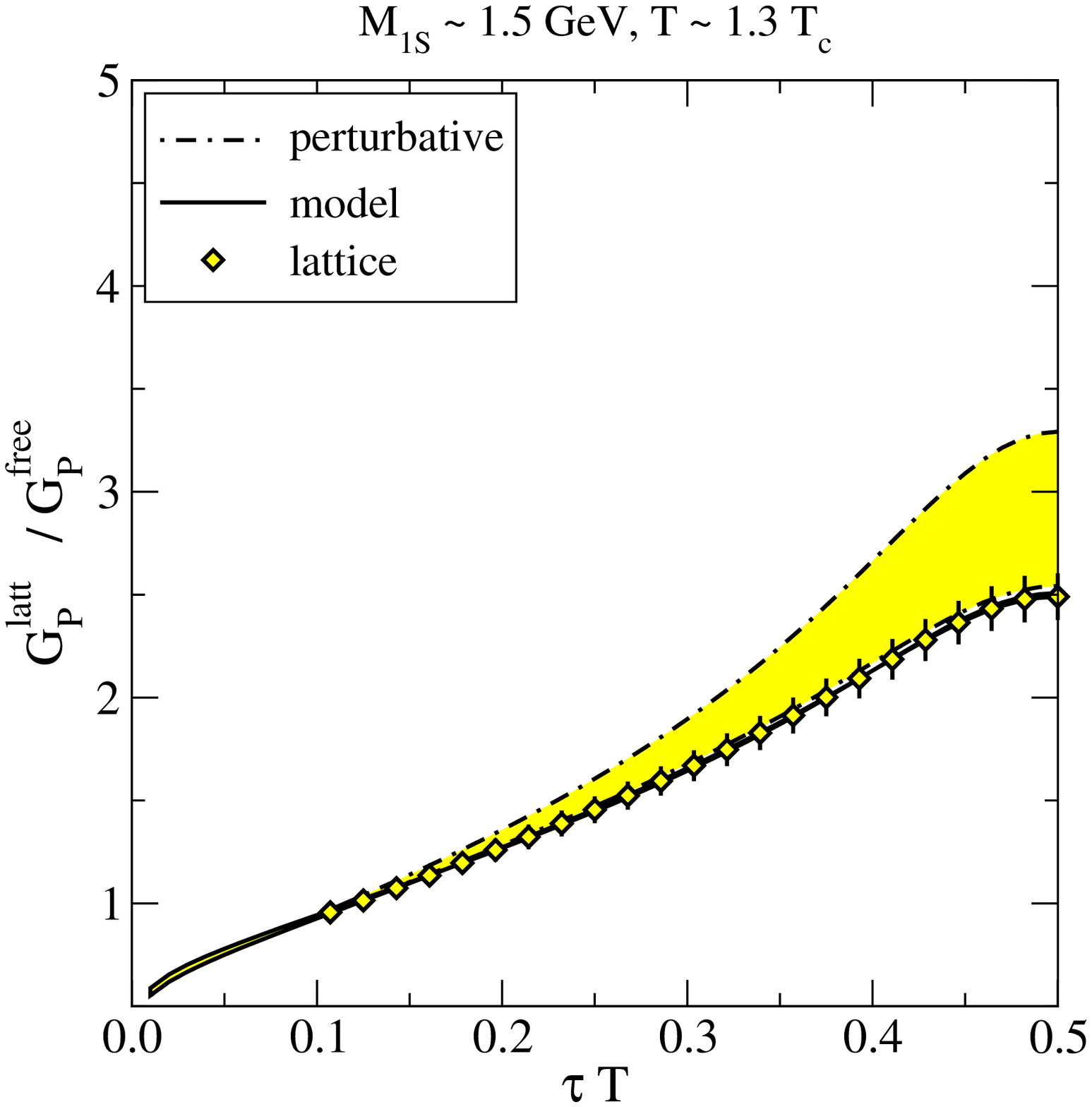}%
}

\vspace*{3mm}

\centerline{%
 \epsfysize=7.5cm\epsfbox{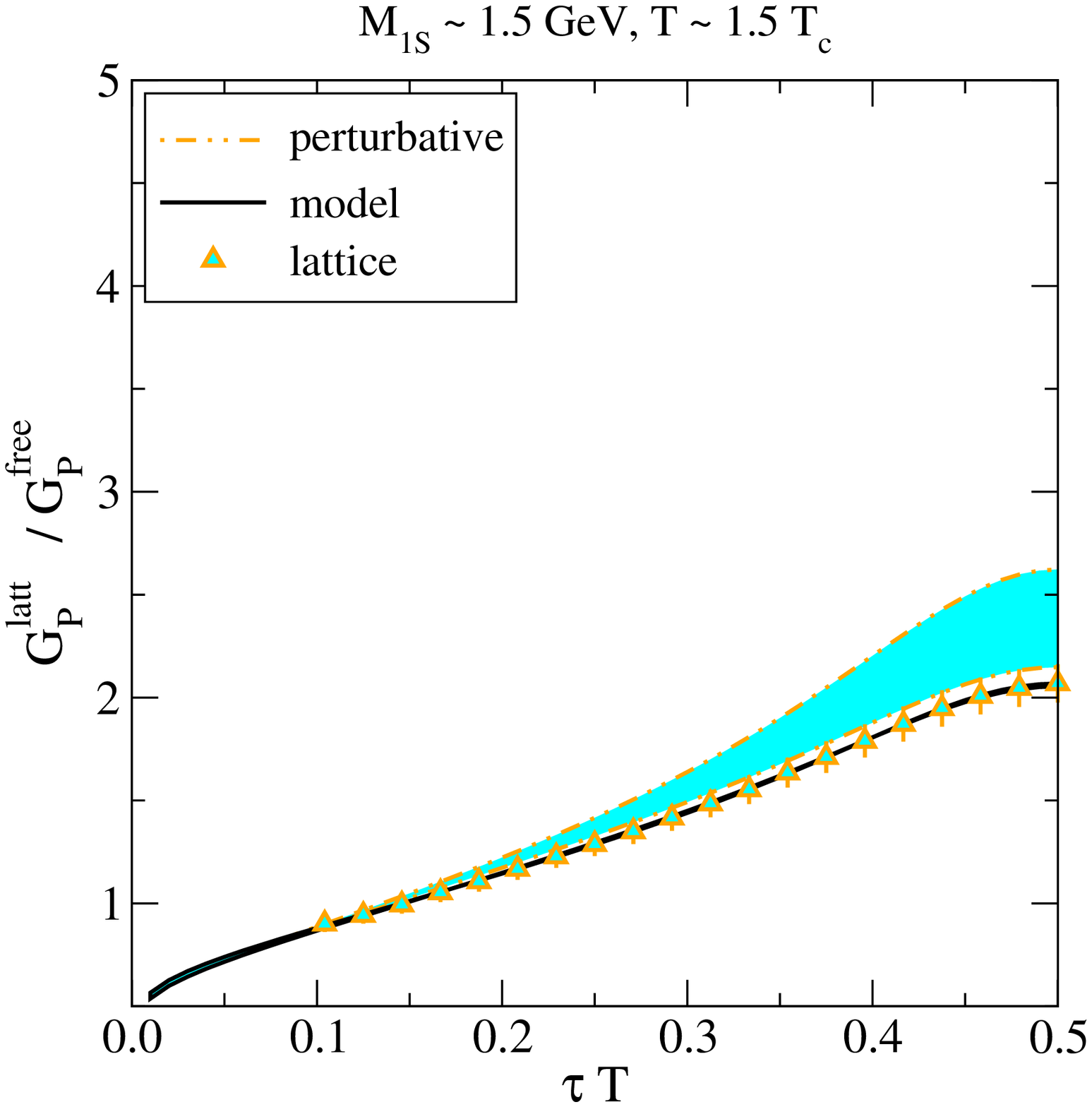}
~~~\epsfysize=7.5cm\epsfbox{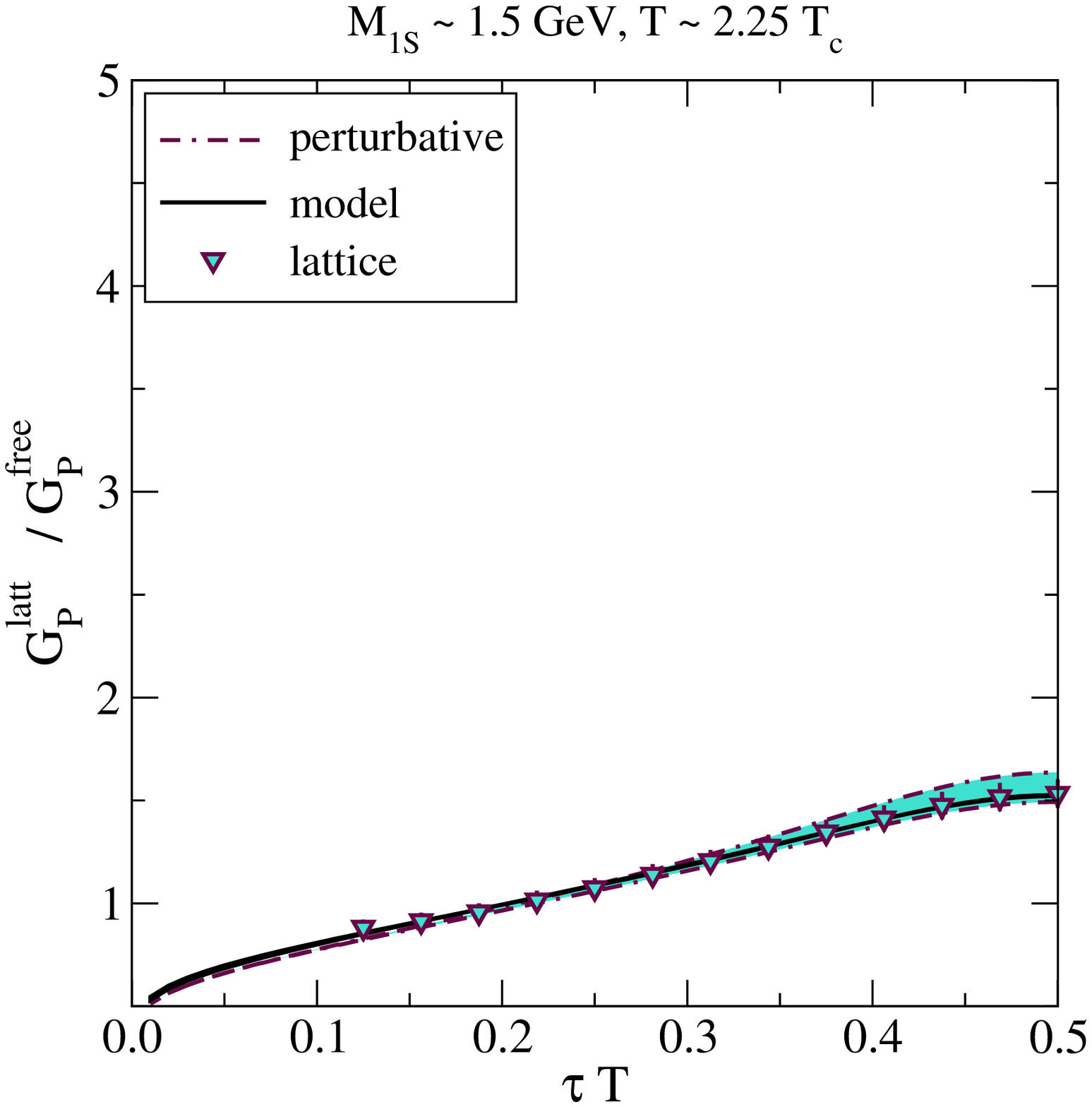}%
}

\caption[a]{\small
Imaginary-time correlators corresponding to 
spectral functions modelled according to \eq\nr{model} 
(narrow bands delineated by solid curves), 
compared with resummed perturbative predictions from
\fig\ref{fig:taudep2}(left) (broad bands) and lattice data from
\fig\ref{fig:taudep2}(right) (open symbols).
}

\la{fig:model2}
\end{figure*}

\begin{figure*}


\centerline{%
 \epsfysize=7.5cm\epsfbox{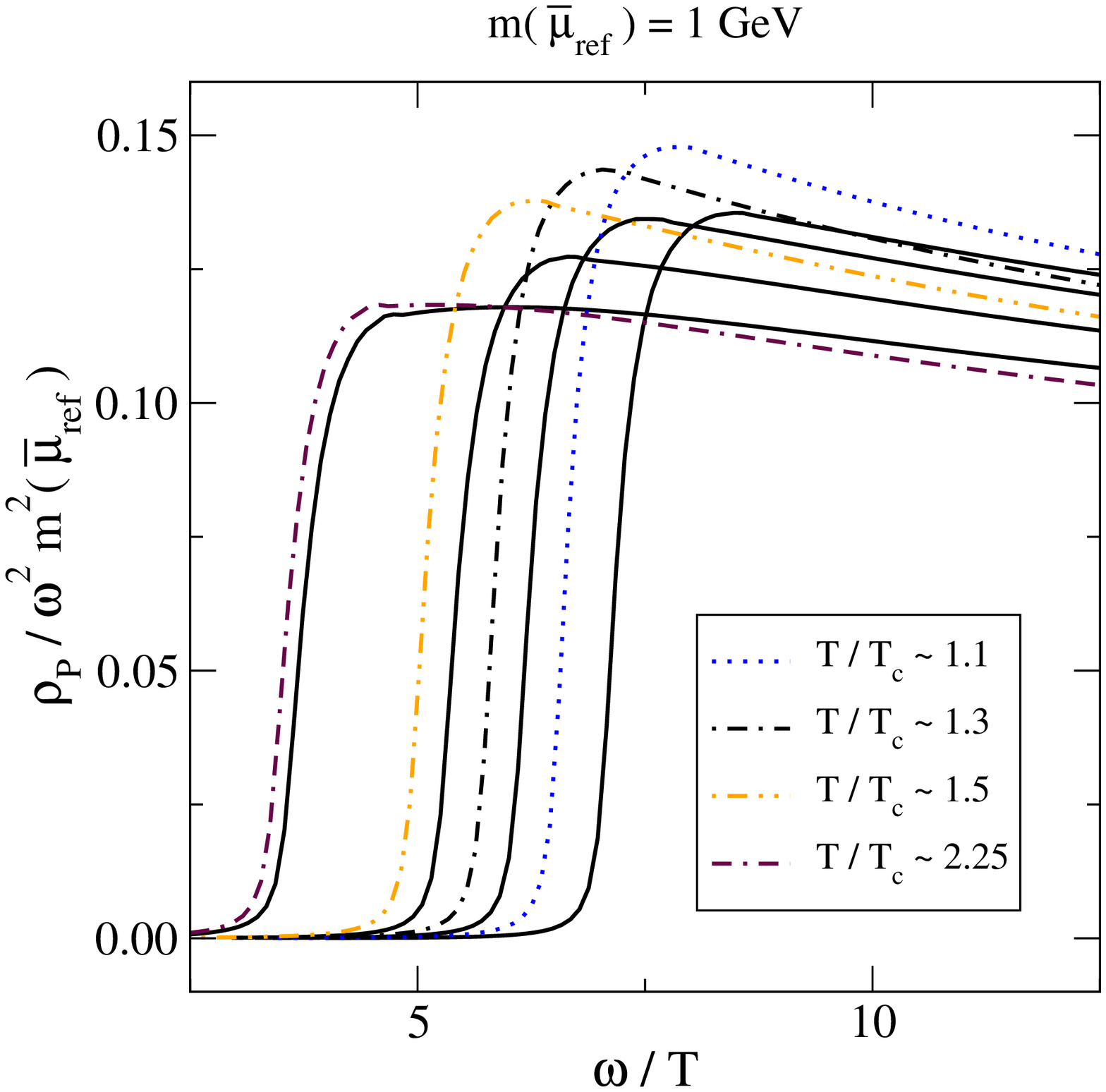}
~~~\epsfysize=7.5cm\epsfbox{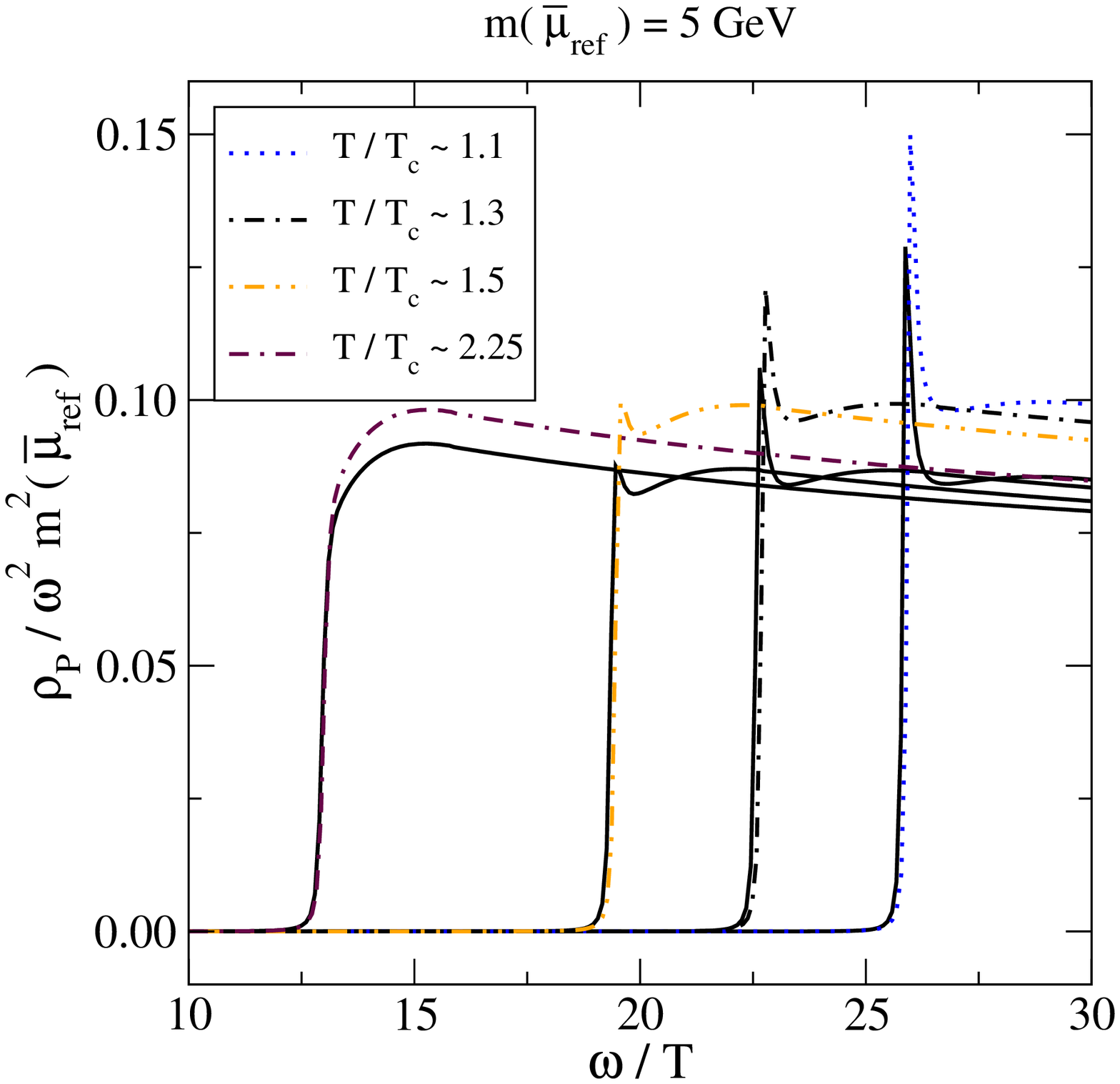}%
}

\caption[a]{\small
Perturbative spectral functions (dash-dotted curves) and their modifications
according to \eq\nr{model} (continuous curves), 
for charmonium (left) and bottomonium (right). 
The continuous curves on the left
yield the imaginary-time correlators labelled by ``model'' in 
\fig\ref{fig:model2}, which agree well with lattice data at all distances. 
These results are elaborated upon in the paragraphs 
following \eq\nr{model}. 
}

\la{fig:model1}
\end{figure*}

There is one clear difference
between the lattice and perturbative sets in \fig\ref{fig:taudep2}: 
the perturbative data display a larger positive ``curvature'' as a 
function of $\tau\Tc$, 
overshooting the lattice data at $\tau\Tc \gsim 0.2$.
 
This difference may have a relatively simple
explanation. In order to demonstrate this, we introduce a ``model'' spectral
function, which is obtained from the perturbative one by adjusting its
overall normalization as well as the threshold location:\footnote{%
 Functional forms which induce a shift only
 at moderate $\omega$ lead to similar results, e.g.\ 
 $ A \, \rho_\rmii{P}^\rmi{pert}
 (\omega - \frac{B}{(\omega^{ }_0 + \omega)^n})$.
 }
\be
 \rho_\rmii{P}^\rmi{model}(\omega) \; \equiv \; 
 A \, \rho_\rmii{P}^\rmi{pert}(\omega - B) 
 \;. \la{model}
\ee
The overall magnitude is adjusted because of uncertainties
related to the perturbative renormalization factors on which
we rely on the lattice side (cf.\ \se\ref{ss:renorm}), 
and the threshold location
because pole-type masses are poorly estimated in  
perturbation theory (cf.\ \se\ref{se:vacuum}).
Results from simple $\chi^2$ fits to the lattice data, 
with $A$ and $B$ treated as fit parameters, are shown in 
\fig\ref{fig:model2}. An excellent representation of the lattice 
data can be found in all cases. 
The same is true, by and large, for the bottomonium case.
The fit parameters are collected in table~\ref{table:fits}.

%
\begin{table}[t]

\small{
\begin{center}
\begin{tabular}{c|ccc|ccc}
 & \multicolumn{3}{c|}{charmonium} 
 & \multicolumn{3}{c}{bottomonium} \\[2mm] 
 $\displaystyle {T } / { \Tc^{ }} $ &
 $\displaystyle A $ &
 $\displaystyle { B } / { T } $ & 
 $\displaystyle {\chi^2} / {\mbox{d.o.f.}} $ &
 $\displaystyle A $ &
 $\displaystyle { B } / { T } $ &  
 $\displaystyle {\chi^2} / {\mbox{d.o.f.}} $ 
 \\[3mm]
 \hline 
  $1.1$ & 1.04 & 0.52 & 0.01  & 0.85 & -0.11 & 0.02  \\ 
  $1.3$ & 1.04 & 0.37 & 0.01  & 0.87 & -0.13 & 0.04  \\
  $1.5$ & 1.02 & 0.33 & 0.02  & 0.87 & -0.11 & 0.10  \\ 
 $2.25$ & 1.06 & 0.16 & 0.08  & 0.93 & -0.04 & 0.28  \\ \hline
\end{tabular} 
\end{center}
}

\vspace*{3mm}

\caption[a]{\small
  Best fit parameters according to \eq\nr{model}. 
  The left set corresponds to charmonium, 
  the right to bottomonium. 
  In these fits the errors of the 
  lattice results at different values of $\tau T$, 
  which are dominated by systematic uncertainties, 
  have been treated as independent of each other. Therefore  
  the results are somewhat qualitative in nature, 
  and we refrain from citing errors. 
 }
\label{table:fits}
\end{table}
%

The model spectral functions corresponding to \fig\ref{fig:model2} 
are shown in \fig\ref{fig:model1}(left), where they are compared with 
the unmodified perturbative ones. The first essential feature is that
some additional spectral weight is needed 
at very large~$\omega$. 
This accounts for the small difference observed
at $\tau T \lsim 0.15$ between the 
lattice and perturbative results. 
This discrepancy may originate from the perturbative renormalization
factors that we have used (cf.\ \se\ref{ss:renorm}). 
Non-perturbative renormalization
would help to clarify this issue. 

The second feature observed in 
\fig\ref{fig:model1}(left) is that the thresholds shift to 
larger masses. This is perfectly admissible, 
given that the procedure we adopted for estimating 
a vacuum ``pole mass'' in \se\ref{se:vacuum} is subject
to large uncertainties, and that thermal mass corrections beyond
the perturbative ones included through \eq\nr{expl}
could be substantial. In fact, it is known from lattice studies
of renormalized Polyakov loop expectation values that close to $\Tc$ 
thermal mass corrections are positive, whereas \eq\nr{expl} 
predicts a negative thermal mass correction, which 
in Polyakov loop measurements 
is observed only at $T \gsim 3\Tc$~\cite{pol}. 

For the bottomonium case, the model spectral functions 
are shown in \ref{fig:model1}(right). The perturbative input
originates from \fig\ref{fig:inter} and contains a resonance 
peak at $T \lsim 1.5 \Tc$. 
At these larger frequencies
non-perturbative mass shifts are barely visible. In contrast the overall
normalization is corrected by a larger amount than for the 
charmonium case, and downwards. 
This difference is not surprising, given that because
of the larger mass, discretization effects, including 
corrections of $\rmO(\alphas^2 a m^{ }_\rmii{L})$, 
imply that our renormalization factors and continuum
extrapolation are likely to contain larger systematic uncertainties. 

We stress that, as \fig\ref{fig:model1}(left) 
shows, no resonance peaks are needed for representing 
the lattice data for the charmonium correlator
even at the lowest temperatures in the
deconfined phase; a modest threshold enhancement is 
perfectly sufficient.
In contrast a thermally broadened resonance peak, 
as predicted by resummed perturbation theory, may well be present 
in the bottomonium case at $T \lsim 1.5\Tc$, 
cf.\ \fig\ref{fig:model1}(right).
 Even though perturbation theory contains inherent uncertainties, such 
 features have been consistently observed in previous 
 computations, however the quantitative properties of the peaks depend
 on the precise approximation under which the thermal potential and
 the gauge coupling have been estimated. 

%
\section{Conclusions and outlook}
\la{se:concl}

We have presented a resummed perturbative estimate of 
the thermal quarkonium pseudoscalar spectral function on one hand 
(\ses\ref{se:vacuum} and \ref{se:thermal}), 
and a continuum-extrapolated quenched lattice measurement of the
corresponding imaginary-time correlator on the other
(\se\ref{se:lattice}). Our main conclusions originate from 
a comparison of these two computations (\fig\ref{fig:model2}). 
Unambiguous but not overwhelming
differences are observed, which call for an interpretation. 

Two possible culprits 
have been put forward in \se\ref{se:model}. 
First, it is plausible
that the perturbative renormalization that we have used for 
the lattice correlators, and other uncertainties 
related to the continuum extrapolation at the very large 
$\beta$-values that we have used, could result in inaccuracies
on a $\sim 5-10$\% level for the overall normalization
of the imaginary-time correlators. This problem is expected to 
be more severe in the bottomonium case, given that 
$\rmO(\alphas^2 a m^{ }_\rmii{L})$ corrections could affect the
continuum extrapolation. Second, the thermal 
quark-antiquark threshold location is not accessible to 
perturbation theory on a quantitative level; treating it 
rather as a fit parameter significantly improves upon the agreement. 
The latter problem is, relatively speaking, more severe in the 
charmonium case, given that the threshold is located at a smaller energy. 

In contrast, there is {\em no need} to modify the perturbative 
charmonium spectral function through resonance peaks at any of
the deconfined temperatures that we have considered, which is 
consistent with a rapid dissociation of the $\eta^{ }_c$ meson
in the deconfined phase of quenched QCD. In the bottomonium case, 
a thermally broadened $\eta^{ }_b$ peak can persist up to $\sim 1.5 \Tc$.
To rephrase these observations more strictly, our statement is that 
perturbative spectral functions, which display these features, 
yield imaginary-time correlators  
perfectly compatible with lattice data, apart from a modest
shift of the threshold location. 
 At the same time the existence of some features beyond the perturbative
 ones cannot be excluded.

In order to consolidate this tentative picture, several future steps 
are needed. To remove the uncertainty concerning overall normalization, 
non-perturbative renormalization is desirable, including the determination
of quark mass effects of $\rmO(a m^{ }_\rmii{L})$ so that 
$\rmO(a)$ improvement is complete~\cite{bali}. Ambiguities originating
from the fact that
our low-temperature runs, needed for scale setting, 
are frozen to the trivial topological sector, 
should be addressed, for instance by taking volume
averages in very large volumes~\cite{largeV}. For insight on 
dissociation patterns, spectral reconstruction 
techniques could be applied to the continuum-extrapolated data. 

We end with an important comment 
concerning the quenched approximation on which this study
was based. Generally, for a comparative value of $T/\Tc$, we would 
expect the quenched deconfined phase to be better described 
by perturbation theory (i.e.\ more weakly coupled)
than the unquenched one. A physics argument
is that in the quenched theory hadrons (glueballs) are heavy, with
$m^{ }_{0_{++}} \gg 1$~GeV, so that the system needs to be
heated up to a high temperature to ``dissolve'' them. Concretely, 
$\Tc \simeq 1.24 \Lambdamsbar > \Lambdamsbar$, 
so that $\alphas^\rmii{(EQCD)} |^{ }_{T\simeq \Tc} \simeq 0.2$
is reasonably ``small''~\cite{gE2}. In contrast, 
the lightest excitations of the unquenched theory are pions, 
with $m^{ }_{\pi} \ll 1$~GeV, and only a modest heating is 
needed to reach the transition temperature (or crossover). Concretely, 
$\Tc \simeq 0.45 \Lambdamsbar < \Lambdamsbar$, 
so that $\alphas^\rmii{(EQCD)} |^{ }_{T\simeq \Tc} > 0.3$ is
undoubtedly too large for perturbation theory 
to apply. 

To summarize, the ``effective'' strong coupling,  
describing soft thermal physics near thres\-hold, 
is likely to be larger in the unquenched theory, and therefore 
the physics of $\eta^{ }_c$ could differ from that in 
the quenched case. Indeed, according to an up-to-date 
potential model study~\cite{bkr1}, the $\eta^{ }_c$ 
is expected to display a resonance peak at $T > \Tc$
in the unquenched theory. 
Therefore our analysis should be repeated for that situation
(which has previously been addressed with relativistic lattice techniques 
in ref.~\cite{bw}, however without a continuum
extrapolation or a comparison with resummed perturbation theory). 
We hope that the present paper helps 
to trace out a path for tackling this ambitious challenge. 

%
\section*{Acknowledgements}

We thank A.~Francis for collaboration at initial stages of this work. 
M.L.\ thanks J.~Piclum for useful discussions several years ago. 
This work was partly supported by 
the German Bundesministerium f\"ur Bildung und Forschung (BMBF) 
under grant 05P15PBCAA, 
by the German Academic Exchange Service (DAAD)
under grant 56268409, 
by the Deutsche Forschungsgemeinschaft (DFG) 
under grant CRC-TR 211 ``Strong-interaction matter under extreme conditions'',
by the National Natural Science Foundation of China 
under grants 11535012 and 11521064, 
by the Swiss National Science Foundation (SNF) 
under grants 200020-168988 and PZ00P2-142524, 
and by the COST Action CA15213 THOR.
Simulations were
performed using JARA-HPC resources at RWTH Aachen and JSC J\"ulich
(projects JARA0039 and JARA0108), 
the OCuLUS Cluster at the Paderborn Center for Parallel
Computing, and the Bielefeld GPU cluster.

%
\appendix
\renewcommand{\thesection}{Appendix~\Alph{section}}
\renewcommand{\thesubsection}{\Alph{section}.\arabic{subsection}}
\renewcommand{\theequation}{\Alph{section}.\arabic{equation}}

%
\section{Strict NLO imaginary-time correlator for any $M/T$}

In this appendix we compute the imaginary-time correlator
$G^{ }_\rmii{P}(\tau)$ up to NLO in strict perturbation theory. 
The purpose is to demonstrate
that it contains no constant contribution, unlike the
corresponding correlators in the vector~\cite{GVtau} 
and scalar~\cite{GStau} channels. 

Making use of the notation
\ba
 \mathcal{J}^{m_1m_2}_{n_1n_2} (\tau) & \equiv & 
 T \sum_{\omega_n} e^{-i \omega_n \tau}
 \left. 
 \Tint{\{P\}}
  \frac{(M^2)^{m_1} (K^2)^{m_2} }
 {
 \Delta^{n_1}_P \Delta^{n_2}_{P-K}
 }
 \right|_{K \,\equiv\, (\omega_n,\vec{0})}
 \;, \la{def_J}
 \\
 \mathcal{I}^{m_1m_2m_3}_{n_1n_2n_3n_4n_5} (\tau) & \equiv & 
 T \sum_{\omega_n} e^{-i \omega_n \tau}
 \left. 
 \Tint{\{P\}Q}
  \frac{(M^2)^{m_1} (K^2)^{m_2}(2Q\cdot K)^{m_3}}
 {(Q^2)_{ }^{\raise0.4ex\hbox{$\scriptstyle n_1$}} 
 \Delta^{n_2}_P \Delta^{n_3}_{P-Q}\Delta^{n_4}_{P-K}
 \Delta^{n_5}_{P-Q-K}}
 \right|_{K \,\equiv\, (\omega_n,\vec{0})}
 \la{def_I}
 \hspace*{-5mm},  \hspace*{5mm}
\ea
where 
$
 \Delta^{ }_P \equiv P^2 + M^2
$
and Matsubara sum-integrals are denoted by 
$\Tinti{ \{ P \} } \equiv T\sum_{ \{ p_n \} } \int_\vec{p}$, 
with $\{ P \}$ denoting fermionic Matsubara momenta,
the tree-level correlator reads
\ba
 G^\rmii{LO}_\rmii{P}(\tau)
 & = & 
 2 \CA M^2 \,
 \Bigl\{
   2 \mathcal{J}^{00}_{10} 
 - \mathcal{J}^{01}_{11}
 \Bigr\}
 \;. \la{GPS_lo} 
\ea
The counterterm contribution ($M_\rmii{B}^2 = M^2 + \delta M^2$) is 
\ba
 \delta G^\rmii{NLO}_\rmii{P}(\tau)
 & = & 2 \CA\, \delta M^2 \,
 \Bigl\{ 
   2 \mathcal{J}^{00}_{10} 
  - \mathcal{J}^{01}_{11}
  - 2 \mathcal{J}^{10}_{20} 
  + 2 \mathcal{J}^{11}_{21}
 \Bigr\}
 \;. \hspace*{5mm} \la{GPS_ct}
\ea
Here
$
  \delta M^2  =  
 - \frac{6 g^2 \CF M^2}{(4\pi)^2} 
 \bigl( \frac{1}{\epsilon} + \ln\frac{\bmu^2}{M^2} + \fr43 + \delta \bigr) 
$, 
where $\delta$ specifies the scheme. 
The 2-loop graphs amount to
(in naive dimensional regularization in $D = 4-2\epsilon$
spacetime dimensions)
\ba
 G^\rmii{NLO}_\rmii{P}(\tau)
 & = & \, 4 g^2 \CA \CF M^2 
 \Bigl\{ \,
   2(1-\epsilon) \, 
     \Bigl[ 
       \mathcal{I}^{000}_{02100}
      - \mathcal{I}^{000}_{12000}
      + \mathcal{I}^{010}_{12010}
      - \mathcal{I}^{010}_{02110}
      + \mathcal{I}^{001}_{11110}
     \Bigr]
   + \mathcal{I}^{020}_{11111}
 \nn 
  & + & 
     2 \, 
       \Bigl[ 
        \mathcal{I}^{000}_{11100}
       + \mathcal{I}^{110}_{11111}
       \Bigr]
   + 4 \, 
       \Bigl[ 
        \mathcal{I}^{110}_{12110}
       - \mathcal{I}^{100}_{12100}
       - \mathcal{I}^{010}_{11110}
       \Bigr]
    + \epsilon
   \,         \mathcal{I}^{010}_{01111}
 \Bigr\} \,
 \;. \la{GPS_nlo}
\ea

All the Matsubara sums appearing in \eqs\nr{GPS_lo}--\nr{GPS_nlo} 
can be carried out. To display the subsequent results, 
we employ the notation of ref.~\cite{GVtau}. 
The LO result becomes
\ba
  G_\rmii{P}^\rmii{LO}(\tau) 
 & = & 
 2 \CA M^2 
 \int_{p} 
  D^{ }_{2E_p}(\tau) 
  \;, \la{GPSLOtaul}
\ea
where $D^{ }_{2E_p} \equiv D^{ }_{E_p E_p}$ and 
\be
 D_{E_1 \cdots E_k}^{E_{k+1} \cdots E_n}(\tau) 
 \equiv
 \frac{
 e^{ (E_1 + \cdots + E_k)(\beta - \tau) +
     (E_{k+1} + \cdots + E_n)\tau } 
   + 
 e^{ (E_1 + \cdots + E_k)\tau +
 (E_{k+1} + \cdots + E_n)(\beta - \tau) } 
 }{[e^{\beta E_1} \pm 1] \cdots [e^{\beta E_n} \pm 1]}
 \;. \la{D_def}
\ee
The sign in the denominator is chosen according to whether the 
particle is a boson ($\epsilon^{ }_q\Rightarrow -$) or a fermion 
($E^{ }_{p}\Rightarrow +$). 
Scheme dependence can be expressed as 
\ba
 \Delta^{ }_\delta G_\rmii{P}^\rmii{NLO}(\tau) 
 & = & 
  - \frac{3 g^2 \CA \CF M^2\, \delta}{4 \pi^2}
 \int_{p} \,
  \biggl( 1 - \frac{M^2 }{2 p^2}   \biggr)  
 \,  D^{ }_{2E_p}(\tau)  
  \;, \la{GPS_LO_tau_delta}
\ea
where $\delta$ was defined below \eq\nr{GPS_ct}. 
Inserting the expressions listed in appendix~A of ref.~\cite{GVtau} for 
$ \mathcal{J}^{m_1m_2}_{n_1n_2} $ and 
$ \mathcal{I}^{m_1m_2m_3}_{n_1n_2n_3n_4n_5} $, 
the remaining NLO contribution reads
\ba
 & & \hspace*{-1cm}
 \frac{ 
    G_\rmii{P}^\rmii{NLO} (\tau)
  }{ 
  4 g^2 \CA \CF M^2
  } = 
 - 
 \int_{p}  
 \frac{D^{ }_{2E_p}(\tau)}{8\pi^2} 
 \nn
 & + & \!\!  \int_{p,q} \!\! \mathbbm{P} \Biggl\{   
 \int_{z} 
 \frac{ [\, D^{ }_{\epsilon_q E_p E_{pq}}(\tau) + 
          D^{\epsilon_q}_{E_p E_{pq}}(\tau)
        \, ] \, M^2}{\epsilon_q E_p E_{pq}
 \Delta_{+-}\Delta_{-+} }
 \biggl[ 
  \frac{\epsilon_q^2 + (E_p + E_{pq})^2 }{4 M^2}
 - \frac{\epsilon_q^2}{\Delta^{ }_{+-}\Delta^{ }_{-+}}
 \biggr]
 \nn 
 &  & \quad - \, 
 \int_{z} 
 \frac{2 D^{E_p}_{\epsilon_q E_{pq}}(\tau)\, M^2}{\epsilon_q E_p E_{pq}
  \Delta_{++}\Delta_{--}
 }
 \biggl[ 
   \frac{\epsilon_q^2 +  (E_p -  E_{pq})^2}{4M^2}
 - \frac{\epsilon_q^2}{\Delta^{ }_{++}\Delta^{ }_{--}}
 \biggr]
 \nn 
 & + &  \frac{ D^{ }_{2E_p}(\tau)}{ 2\epsilon_q^3} 
  \, \biggl[ \;
 1 + \frac{
    E_p^2 (E_{pq}^+ - E_{pq}^-)
     -
    p\epsilon_q (E_{pq}^+ + E_{pq}^-)
    }{2p(\epsilon_q^2 - E_p^2 )}
 + \frac{\epsilon_q^2 M^2 (E_{pq}^+ - E_{pq}^-) }
    {p ( \epsilon_q^2 - E_p^2 ) E_{pq}^+  E_{pq}^- }
 \nn &   & 
 \quad + \,
 \frac{ 2 E_p^2  - M^2 }{2p E_p}
 \biggl(  
     \ln\biggl| \frac{(E_p - p)(2p + \epsilon_q)}
                     {(E_p + p)(2p - \epsilon_q)}
        \biggr|
    +\ln\biggl| \frac{1 - {\epsilon_q^2} / {(E_p + E_{pq}^+)^2}}
                 {1 - {\epsilon_q^2} / {(E_p + E_{pq}^-)^2}}
        \biggr|
 \biggr) \;
 \biggr]
 \nn 
 &  + &
 \frac{D^{ }_{2E_p}(\tau)  \nB{}(\epsilon_q)}{\epsilon_q}
  \biggl[ 
    \frac{1}{\epsilon_q^2} - 
  \frac{1}{2 p^2} +  \frac{\epsilon_q^2+ 2 E_p^2  - M^2 }{2 p E_p \epsilon_q^2}
   \ln \biggl(\frac{E_p - p}{E_p + p}
   \biggr) 
 \; \biggr]
 \nn &  + &    
 \frac{D^{ }_{2E_p}(\tau)  \nF{}(E_{q})}{E_{q}} \, 
 \biggl[ \;
  \frac{E_q^2 - 3 E_p^2 + M^2}{2 p^2 ({p^2-q^2})}
 \nn &   &  \quad + \, 
 \frac
 {q^2 + E_p^2 }
 {2 p q (E_p - E_q) E_p}
 \biggl( 
   \ln \biggl| \frac{p+q}{p-q} \biggr| + 
  \frac{E_q}{E_p + E_q} 
   \ln \biggl| \frac{M^2 + E_p E_q + p q}{M^2 + E_p E_q - p q} \biggr|
 \biggr)
 \biggr]
 \Biggr\} 
 \;,  \la{GPS_NLO_tau_final}
\ea
where $\mathbbm{P}$ denotes a principal value; 
$\nB{}$ and $\nF{}$ are Bose and Fermi distributions; 
$\epsilon^{ }_q \equiv q$; 
$
 E^{ }_p \equiv \sqrt{p^2 + M^2}
$; 
$
 E^{ }_{pq} \equiv \sqrt{p^2 + q^2 + 2 p q z + M^2}
$;
$ 
 E^{\pm}_{pq} \equiv  E^{ }_{pq} |^{ }_{z=\pm}
$;  
$
  \Delta^{ }_{\sigma\tau} \equiv \epsilon^{ }_q 
 + \sigma E^{ }_p + \tau E^{ }_{pq}
$.

It is observed from 
\eqs\nr{GPSLOtaul}--\nr{GPS_NLO_tau_final}
that all terms contain non-trivial $\tau$-dependence, 
i.e.\ that there is no constant contribution
in $G^{ }_\rmii{P}(\tau)$ at NLO. 

%
\section{Strict NLO spectral function for $M/T \gg 1$}

We specify here the NLO expression for the thermal pseudoscalar spectral
function, up to corrections suppressed by $e^{-M/T}$
(techniques for including these are discussed in ref.~\cite{fullM}). 
The main goal is to demonstrate how the spectral function behaves 
below the threshold, i.e.\ for $2 M - \omega \gg \alphas^2 M$, 
a regime that cannot be addressed with the methods of \se\ref{se:thermal}. 
Physically, the NLO corrections originate from virtual 
heavy quark self-energy and gluon exchange 
contributions, as well as from 
real gluon emissions and absorptions. 

The spectral functions corresponding to the master
sum-integrals appearing in \eq\nr{GPS_nlo} have been worked
out in ref.~\cite{nlo}.
Making use of the objects
$S^i_j(\omega)$ defined there, the spectral
function can be written as 
\ba
 \frac{\left. \rho^{ }_\rmii{P}(\omega) \right|_\rmi{raw}}{M^2} & \equiv & 
 2 \CA\, \omega^2 S^{ }_1(\omega)
 \nn 
 & + & 4 g^2 \CA \CF 
 \biggl\{ 
   - \frac{3}{(4\pi)^2}
   \biggl(
     \frac{1}{\epsilon} + \ln\frac{\bmu^2}{M^2} + \fr43 + \delta
   \biggr) 
 \, \omega^2 S^{ }_1(\omega)
 \nn[1mm] 
 & & + \;  
 \biggl[ 
   \frac{6 M^2}{(4\pi)^2}
   \biggl(
     \frac{1}{\epsilon} + \ln\frac{\bmu^2}{M^2} + \fr43 + \delta
   \biggr)
 - \frac{T^2}{6}  
 \biggr]\, \omega^2 S^{ }_2(\omega)
 \nn[1mm] 
 & & +\; 4 \omega^2 
 \bigl[ S_4^0(\omega) 
  - M^2 S_5^0(\omega)
 \bigr]
 + 2 (1-\epsilon) 
 \bigl[ 2 S_4^1(\omega)
   +  \omega^2 S_5^2(\omega)
 \bigr]
 \nn  & &
 +\; (\omega^2 - 2 M^2)\,\omega^2 S_6^0(\omega) 
 - \epsilon\, \omega^2 S_6^2(\omega)
 \biggr\} + \rmO(g^4) \;.
   \la{full_raw_PS}
\ea
We have set here $\epsilon\to 0$ whenever the master sum-integral
that it multiplies is finite. Expressing the 
result as a sum of a vacuum and thermal part, 
\be
 \left.
 \rho^{ }_\rmii{P}(\omega) \right|_\rmi{raw}
 \; = \; 
 \left. 
 \rho^{ }_\rmii{P}(\omega) \right|^\rmi{vac} 
 \; + \; 
 \left. 
 \rho^{ }_\rmii{P}(\omega) \right|^\rmii{$T$} 
 \;,
\ee
where the vacuum part is {\em defined} to include 
also the explicit $T^2$ visible in \eq\nr{full_raw_PS}, 
and inserting the expressions for the functions 
$S^i_j(\omega)$ from ref.~\cite{nlo},  the vacuum part reads
\ba
 \frac{ \left. \rho^{ }_\rmii{P}(\omega) \right|^\rmi{vac} }{M^2}
 & \equiv & 
  \theta(\omega - 2 M)\,   
  \frac{\CA\, \omega (\omega^2 - 4 M^2)^{\fr12}}{8\pi}
 \nn 
 & + &  \theta(\omega - 2 M)\, 
 \frac{4 g^2 \CA \CF }{(4\pi)^3 } 
 \biggl\{  
   (\omega^2 - 2 M^2) 
 L^{ }_2 \biggl( \frac{\omega - \sqrt{\omega^2 - 4 M^2}}
 {\omega + \sqrt{\omega^2 - 4 M^2}} \biggr)
 \nn &  & 
 \; + \, 
 \biggl(\frac{3 \omega^2}{2}  - 2 M^2  + \frac{ 3 M^4}{\omega^2} \biggr)
  \,\mathrm{acosh} \biggl( \frac{\omega}{2 M} \biggr)
 \nn &  & 
 \; - \, \omega (\omega^2 - 4 M^2)^{\fr12}
 \biggl[  
  \ln  \frac{\omega (\omega^2 - 4 M^2)}{M^3}
  -\fr98  - \frac{3 M^2}{4\omega^2} 
 \biggr]
 \nn &  & 
 \; - \, 
 \frac{\omega}{24 (\omega^2 - 4 M^2)^{\fr12} }
 \biggl[
   18 (\omega^2 - 6 M^2) \delta + (4\pi)^2 T^2 
 \biggr]
 \biggr\} + \rmO(g^4) \;.
   \la{full_vacS}
\ea
Here the function $L^{ }_2$ is defined as~\cite{jf}
\be
 L^{ }_2(x) \; \equiv \; 
 4 \, \mathrm{Li}_2 (x) + 2 \, \mathrm{Li}_2(-x)
 + [2 \ln(1-x) + \ln(1+x)] \ln x
 \;. \la{L2} 
\ee

The last term in \eq\nr{full_vacS} is divergent 
at the threshold $\omega \sim 2 M$. 
There is a unique choice which avoids this
at all $T$, namely adopting the pole mass
scheme ($\delta \equiv 0$) and resumming the explicit thermal correction
into an effective mass modifying the LO result~\cite{dhr}, 
\be
 M_\rmi{eff}^2 \equiv M^2 + \frac{g^2 T^2 \CF}{6}
 \;. \la{Meff}
\ee
We then re-interpret the mass appearing in 
$
 {\left. \rho^{ }_\rmii{P}(\omega)
 \right|^\rmi{vac}} 
 / {M^2} 
$ 
as the thermal effective mass. 
The correction in \eq\nr{Meff} is very small in practice
for $T \ll M$,  
$
 M^{ }_\rmi{eff} - M  \sim \alphas T^2 / M 
$; 
in particular it is smaller than the thermal mass originating
from Debye screening in \eq\nr{expl}, which 
is of order 
$ 
  - \alphas^{3/2} T
$. 

Consider finally 
$
 \left. 
 \rho^{ }_\rmii{P}(\omega) \right|^\rmii{$T$} 
$.
Going over to the notation of \eq\nr{def_v}, it can be represented as
\ba
 \left. R^p_\rmi{c}(\omega) \right|^\rmii{$T$} \!\!\! & = & \!\!\!
 \frac{2 \alphas \CF}{\pi \omega^2}
 \int_0^\infty \! \dd q \, \frac{\nB{}(q)}{q} 
 \biggl\{
 \nn  
 & & \hspace*{-0.1cm}
   \theta(\omega) \, 
   \theta\Bigl(q- 
    \frac{4M^2-\omega^2}{2\omega}
    \Bigr)
   \biggl[ 
  -\;    \sqrt{\omega(\omega+2 q)}
    \sqrt{\omega(\omega+2 q)- 4M^2}
  \nn & & \hspace*{0.1cm}
  +\; 2 \Bigl(q^2 + (\omega + q)^2 - 2 M^2 \Bigr)
  \, \mathrm{acosh} \sqrt{\frac{\omega(\omega+2q)}{4M^2}} 
   \biggr]
   \nn & + & \hspace*{-0.1cm}
   \theta(\omega - 2M) \, 
   \theta\Bigl(\frac{\omega^2-4M^2}{2\omega}-q\Bigr)
   \biggl[ 
   - \;   \sqrt{\omega(\omega-2 q)}
    \sqrt{\omega(\omega-2 q)- 4M^2}
  \nn & & \hspace*{0.1cm}
   +\;  2 \Bigl( q^2 + (\omega - q)^2 - 2 M^2 \Bigr)
  \, \mathrm{acosh} \sqrt{\frac{\omega(\omega-2q)}{4M^2}} 
   \biggr]
   \nn & + & \hspace*{-0.1cm}
  \theta(\omega - 2M) \biggl[ 
   2 \, \omega \sqrt{\omega^2 - 4 M^2}
  \nn & & \hspace*{0.1cm}
   -\; 4 \Bigl(\omega^2 + 2 q^2 - 2 M^2 \Bigr)
   \, \mathrm{acosh} \biggl( \frac{\omega}{2M} \biggr) 
  \biggr] \biggr\}
 + \rmO(e^{-M/T},\alphas^2)
 \;. \la{full_TS}
\ea
This is finite and small around the threshold:
$
 \lim_{\omega\to 2 M}
 \left. R^p_\rmi{c}(\omega) \right|^\rmii{$T$} \sim \alphas \CF (T/M)^{3/2}
$. 

The main importance of \eq\nr{full_TS} lies in 
its first term,  which 
remains non-zero below the threshold ($\omega < 2 M$)
and gives the dominant
contribution in this regime.
Writing $\omega = 2 M + E'$ and assuming $|E'| \ll M$, the 
restriction on the integration variable $q$ reads $\theta(q + E')$. 
Therefore, for $E' < 0$, the result is 
$\sim \alphas\CF (T/M)^{3/2} \nB{}(|E'|) 
 \sim \alphas\CF (T/M)^{3/2} e^{-|E'|/T}$.

\small{
%

}

\end{document}